\documentclass[11pt]{article}
\usepackage{amsmath,  amssymb}
\usepackage{graphicx}
\begin{document}

\title{Evolutionarily Stable Sets in Quantum Penny Flip Games}
\author{Tina Yu and Radel Ben-Av }
\date{}
\maketitle

\begin{abstract}
In game theory, an Evolutionarily Stable Set (ES set) is a set of Nash Equilibrium (NE) strategies that give the same payoffs. Similar to an Evolutionarily Stable Strategy (ES strategy), an ES set is also a \emph{strict} NE. 
This work investigates the evolutionary stability of classical and quantum strategies in the quantum penny flip games.
In particular, we developed an evolutionary game theory model to conduct a series of simulations where a population of mixed classical strategies from the ES set of the game were invaded by quantum strategies. We found that when only one of the two players' mixed classical strategies were invaded, the results were different. In one case, due to the interference phenomenon of superposition,  quantum strategies provided more payoff, hence successfully replaced the mixed classical strategies in the ES set. In  the other case, the mixed classical strategies were able to sustain the invasion of quantum strategies and remained in the ES set. Moreover, when both players' mixed classical strategies were invaded by quantum strategies, a new quantum ES set emerged. The strategies in the quantum ES set give both players payoff 0, which is the same as the payoff of the strategies in the mixed classical ES set of this game.
\end{abstract}

\section{Introduction}
Game theory applies probability theory to address uncertainty in the decision making process. In the classical world,  players' decisions are stored in classical channels and the information is processed based on classical mechanics. With the advent of quantum computing, it is natural to ask if quantum mechanics would change the dynamics of game playing. To put it another way, if the decisions are stored in quantum channels,  communicated and processed under quantum mechanics, would a game still produce the same result?

In 1999, Meyer \cite{meyer} explored quantum \textit{superposition} to play a two-person zero-sum game: the quantum penny flip game. 
He reported that due to the \textit{interference} phenomenon of superposition, the game that originally has a Nash Equilibrium (NE) strategy that gives both players equal probability to win in the classical world has become biased:  when the first player is allowed to use quantum strategies, this player always outperforms the other player who uses mixed classical strategies. 
Around the same time, Eisert and colleagues \cite{elsert_etal} investigated quantum \textit{entanglement} in playing the Prisoners' Dilemma (PD) game. They quantized the game by first entangling two qubits using a unitary operator (explained in Section \ref{qip}) that was known to both players. After that, the two qubits were distributed to the two players to encode their decisions. Once returned, the two qubits were disentangled using another unitary operator. Finally, the two qubits were measured and the results were used to calculate their payoffs.  By leveraging quantum entanglement, the authors reported that the dilemma of the game did not exist any more. Moreover, the \emph{mutual-defection} classical NE strategy was replaced by a quantum NE strategies that gave both players higher payoffs. Since then, various new development on quantum games have been reported, such as the quantization\footnote{Quantization here refers to ``deriving a quantum version of a classical algorithm",  which is different from ``the process of converting analog to digital signals" that is more popular in the wider scientific community.} of the Battle of the Sexes game \cite{marinatto_weber}, the Hawk-Dove game \cite{nawaz_toor}, the Monty Hall problem \cite{flitney_abbottt} and evolutionary quantum games \cite{kay_etal}.


Although with higher payoffs, are those quantum NE strategies evolutionarily stable? An NE strategy is evolutionarily stable if a population of players have adapted the strategy, natural selection alone is sufficient to prevent it from being invaded by any mutant strategies  \cite{maynardSmith} (more details in Section \ref{ne_es}).  Iqbal and Toor \cite{iqbal_toor_2001} were the first to investigate this problem using the PD game. They reported that when invaded by a group of {\it one-parameter} quantum strategies, the classical mutual-defection NE strategy remained evolutionarily stable. However, when invaded by a group of {\it two-parameter} quantum strategies, they were no longer stable and were replaced by a higher-payoff quantum \emph{mutual-collaboration} strategy.  Moreover, the new dominating mutual-collaboration quantum strategy was evolutionarily stable against the invasion of any two-parameter quantum strategies. 

The objective of this study is to investigate the evolutionary stability of classical and quantum strategies in the quantum penny flip game, which was introduced by Meyer  \cite{meyer} (see Section \ref{pqgame} for more details). In addition to deriving analytical results, we are also interested in understanding the dynamics of strategy changes during the game. Evolutionary game theory (EGT) models \cite{maynardSmith}  are ideal vehicles to provide such insight. In an EGT model, there is a strategy replication rule, in addition to the game contesting rules. The extra rule specifies how fitter strategies are multiplied and how less fit strategies are culled out of a population. Games are played repeatedly for many generations for the dynamics of the strategy changes in the population to emerge. EGT models have been successfully applied to analyze strategy changes, the success/failure of a strategy and the existence of equilibrium strategies in a game \cite{hofbauer_sigmund}.

This research developed an EGT model to study the quantum penny flip game. In particular, we used the model to conduct a series of computer simulations where a population of mixed classical strategies from the Evolutionarily Stable Set (ES set) of the game were invaded by quantum strategies. The game has two players where one player has single move while the other player has two moves. If both players use classical strategies, the two-move player does not have advantage over the one-move player; the game always ends with an equilibrium state where both players receive the same payoff of 0. However, if the two-move player is allowed to use quantum strategies while the one-move player is not, this is no longer true. Our simulation results show the following interesting phenomena:
\begin{itemize}
\item The two-move player's mixed classical strategies are not evolutionarily stable under the invasion of pure quantum strategies, where one particular kind of quantum strategy gives a higher payoff when playing against the one-move player's mixed classical strategies. This result is identical to that reported in \cite{meyer}. Moreover, once they become the dominating strategies in the population, this kind of quantum strategies  are evolutionarily stable against the invasion of any classical and quantum strategies and make the two-move player win the game with certainty.
\item The one-move player's mixed classical strategies are evolutionarily stable under the invasion of pure quantum strategies when playing against the two-move player's mixed classical strategies. In other words, pure quantum strategies do not provide advantage over classical strategies for the one-move player in this game.
\item When both players' mixed classical strategies were invaded by quantum strategies, a new quantum ES set emerged. The strategies in the quantum ES set give both players payoff 0, which is the same as the payoff of the strategies in the mixed classical ES set of this game.
\end{itemize}

In addition to being the first to investigate evolutionary stability of classical and quantum strategies in the quantum penny flip game, this work also made the following novel contributions:
\begin{itemize}
\item 
It proposed and implemented an EGT model to conduct systematic study of Evolutionarily Stable (ES) strategies for quantum games in general and the quantum penny flip game in particular (see Section \ref{qess}). 
\item 
It analyzed and interpreted the EGT model simulation results for 3 variations of the quantum penny flip games (see Section \ref{pure_quantum} and Section \ref{mixed_quantum}).
\item It identified a quantum ES set where all strategies in the set give both players payoff 0, which is the same as the payoff of the strategies in the mixed classical ES set of this game.
\end{itemize}

This work is significant in the following ways:
\begin{itemize}
\item	It demonstrates how evolutionary game theory models can advance our understanding of quantum game theory.
\item	It demonstrates that even the simple quantum penny flip game has a rich problem landscape, which could be an inspiration for the investigation of other quantum games using a similar approach.
\end{itemize}

The paper is organized as follows. Section \ref{ne_es} first explains NE and ES strategies in two-player symmetric and asymmetric games. After that, their extension to the Evolutionarily Stable Set (ES set) under evolutionary game theory models are described. Section \ref{qip} gives a brief introduction to quantum information processing. In Section \ref{pqgame}, the quantum penny flip game is described  and its mixed classical NE strategies that form the ES set of the game are analyzed. Section \ref{qess} presents our work investigating classical and quantum ES sets in 3 variations of the game. Finally, Section \ref{conclude} concludes the paper and outlines our future work.

\section{Nash Equilibrium \& Evolutionary Stability}\label{ne_es}
In classical game theory, a profile of all players' strategies is a Nash Equilibrium (NE) if none of the players can do better by changing his or her strategy unilaterally. 
It is assumed that all players have knowledge of the other players' strategies and are allowed to use that information to maximize their own payoffs.

In mathematical terms, a strategy profile $x^*=\{x_i^*\}$ is a NE under the following condition:
\begin{equation}
\forall i\in \mathbb{N}, x_i \in S_i, x_i\neq x_i^* : f_i(x_i^*, x_{-i}^*) \geq f_i(x_i, x_{-i}^*)
\end{equation}
where $x_i$ is a strategy of player $i$ within his strategy set $S_i$ and $f_i$ is his payoff function. Also, $x_{-i}^*$ is a strategy profile of all players except for player $i$.

Evolutionary stability is a refinement of NE. A NE strategy is an Evolutionarily Stable (ES) strategy if when it is fixed in a population, no alternative (mutant) strategy can invade the population successfully.  In a two-player \textit{symmetric} game, where both players use identical strategies set and have identical payoff function, $x$ is an ES strategy under the following conditions:
\begin{align}
&f(x, x) > f(y, x), \\
&\textbf{if } f(x, x) = f(y, x) \textbf{ then } f(x, y) > f(y, y)
\end{align}
where  $y$ is any alternative (mutant) strategy and $f$ is the payoff function of the game. 

In two-player \textit{asymmetric} games, one population is not sufficient to model the game. Hofbauer and Sigmund \cite{hofbauer_sigmund}  modeled the game using two populations, where each population contains strategies that belong to one of the two players to play against each other under two different payoff functions. A pair of strategies ($p$, $q$) is an \textit{NE pair} if the following condition holds:

$\forall (x, y) \in S_A \times S_B: $
\begin{align}\label{NE}
&f_A(p, q) \geq f_A(x, q) \textbf{ and }  f_B(p, q) \geq f_B(p, y); 
\end{align}
where $S_A$ and $S_B$ are the strategy sets for populations $A$ and $B$ respectively; $f_A$ and $f_B$ are the payoff functions for strategies in population $A$ and $B$ respectively. 

Moreover, $(p,q)$ is an \textit{ES} pair under the following condition:

$\forall x \neq p, y \neq q: $
\begin{align} \label{asymmetric}
&f_A(p, q) > f_A(x, q) \textbf{ and }  f_B(p, q) > f_B(p, y); 
\end{align}
Equation \ref{asymmetric} is the definition of \textit{strict} NE. Hence, in two-player asymmetric games, only \textit{strict} NE strategies are \textit{ES} strategies.  

Thomas \cite {thomas} extended the concept of ES strategies to \emph{Evolutionarily Stable Set (ES set)} for games which do not have an ES strategy but have a continuum of NE strategies that give the same payoffs. These NE strategies form an ES set, which is a strict NE like an ES strategy for these games. 

Thomas showed that under the standard evolution dynamics, if an EGT model starts with a population that is in the neighborhood of an ES set, the population would converge towards some of the strategies of the ES set. In Section \ref{qess}, we will seed our EGT model with strategies from the mixed classical ES set  and then examine if the population converges to any strategies of an quantum ES set, if it exists.


In order to apply the concept of ES sets to quantum strategies, we have to address one important question: ``does evolution take place in the quantum realm?" More precisely, could selection operate on superpositions without measurement? Could quantum information be mutated and inherited from one generation to another? According to Quantum Darwinism \cite{zurek}, the answer is ``yes": quantum states are selected against each other in favor of a stable pointer state. We will give a brief introduction to quantum information processing in the following section to pave the way for our investigation of quantum ES sets.

\section{Quantum Information Processing} \label{qip}
In classical computing, information is stored in binary 0 or 1. By contrast, quantum information resides in superpositions of 0 and 1. 
Quantum information processing takes place on these superpositions simultaneously using unitary operators (explained later).
During the process, however, no probing is allowed to know the intermediate states of the information. 
This is because once measured, quantum information collapses into classical value of 0 or 1. Meanwhile, all superposition information is destroyed. Therefore, measurement is only performed at the end of the information processing to obtain the final result.

A quantum bit (\emph{qubit}) is the counter-part of a classical bit. Using Dirac notation \cite{dirac}, a qubit is represented as a linear combination of  basis states $|0\rangle$ and $|1\rangle$:
\begin{equation}
|\psi \rangle = a |0\rangle + b |1\rangle = \begin{bmatrix}a \\ b \end{bmatrix}, a, b \in \mathbb{C}  
\end{equation}
The state of a qubit, hence, is a vector in a two-dimensional complex vector space. The states $|0\rangle$ and $|1\rangle$ are computational basis states, which form an orthonormal basis for this vector space. 

When a qubit is measured, it collapses to $|0\rangle$ with probability $a\bar{a}$\footnote{The $\bar{a}$ defines complex conjugate of $a$.}, or to $|1\rangle$ with probability $b\bar{b}$. Hence, $a\bar{a}+b\bar{b}=1$. 

In addition to the above vector representation, the state of a qubit can also be formulated using the density matrix \cite{vonNeumann}:
\begin{equation}
\rho=|\psi\rangle\langle\psi|
\end{equation}
For example, the associated density matrix for $|\psi\rangle=|0\rangle$ is: 
\begin{equation}
\rho=\begin{bmatrix}1 \\ 0 \end{bmatrix}\begin{bmatrix}1&0 \end{bmatrix}=\begin{bmatrix} 1 & 0 \\ 0 & 0 \end{bmatrix}.
\label{classical}
\end{equation}
Similarly, the associated density matrix for $|\psi\rangle=\frac{1}{\sqrt{2}}(|0\rangle+|1\rangle)$ is: 
\begin{equation}
\rho=\begin{bmatrix}\frac{1}{\sqrt{2}} \\ \frac{1}{\sqrt{2}} \end{bmatrix}\begin{bmatrix} \frac{1}{\sqrt{2}}&\frac{1}{\sqrt{2}} \end{bmatrix}=\begin{bmatrix} \frac{1}{2} & \frac{1}{2} \\ \frac{1}{2} & \frac{1}{2} \end{bmatrix}.
\label{quantum}
\end{equation}
A mixed quantum state is a linear combination of pure state $|\psi_i\rangle$, each with probability $p_i$:
\begin{equation}
\rho=\sum{p_i}{|\psi_i\rangle\langle\psi_i|}. 
\end{equation}
For example, the associated density matrix for a mixed quantum state being in $|0\rangle$ and $|1\rangle$ with equal probability is: 
\begin{equation}
\rho=\frac{1}{2}\begin{bmatrix}1 \\ 0 \end{bmatrix}\begin{bmatrix}1&0 \end{bmatrix}+\frac{1}{2}\begin{bmatrix} 0 \\ 1 \end{bmatrix}\begin{bmatrix}0&1 \end{bmatrix}=\begin{bmatrix} \frac{1}{2} & 0 \\ 0 & \frac{1}{2} \end{bmatrix}.
\label{mixed2}
\end{equation}
The transformation of a quantum state to another is through a unitary operator $U$, where $UU^\dagger\footnote{The $\dagger$ notion defines Hermitian conjugate.}=I$:
\begin{equation}
\rho_{i+1}=U\rho_{i}U^\dagger
\end{equation}
A quantum state can also be transformed by a mix of $j$ unitary operators, each with probability $p_j$:
\begin{equation}
\rho_{i+1}=\sum {p_j}U_j\rho_{i}U_j^\dagger
\end{equation}
The trace of a density matrix, $\rho$, is always 1. The top left corner value, $\rho$(1,1), gives the probability of the qubit being measured as $|0\rangle$. The bottom right corner value, $\rho$(2,2), gives the probability of the qubit being measured as $|1\rangle$.

For example, the measurement of $\rho$ in Equation (\ref{classical}) produces classical value 0, while the measurement of $\rho$ in Equations (\ref{quantum}) \& (\ref{mixed2}) produces classical value of 0 and 1 with equal probability. From the measurement point of view, the two states in Equations  (\ref{quantum}) \& (\ref{mixed2}) are identical. However, if $\rho$ is an intermediate state, not to be measured, farther transformation of the two $\rho$ may lead to different states, hence different final measurements. 
The $\rho$ in Equation (\ref{quantum}) describes a quantum state as a linear superposition, where \textit{quantum interference} may cancel or enhance the probability of the measurement results. By contrast, the $\rho$ in Equation (\ref{mixed2}) describes a  classical state, where the interference phenomenon does no exist. 
  
The above quantum states and unitary operations can be extended to multiple qubits using the tensor operator $\otimes$.
However, since the quantum penny flip game only operates on one qubit, we will not discuss multiple qubits operations. We refer interested readers to \cite{nielsenChuang}.

\section{Quantum Penny Flip Games}\label{pqgame}
Meyer \cite{meyer} quantized the classical Matching Pennies game in the following ways. 
Initially, the penny is placed in a closed box head-up.
Next, the first player (Q) is allowed to flip the penny. 
Next, the second player (Picard) is allowed to do the same.
After that, Q has a second turn to flip the penny if he wishes.
Since the penny flipping is carried out in a closed box, the intermediate state of the penny is unknown. 
It is only after Q has made his second move, the box is opened and the state of the penny is measured.
If the penny is head-up, Q wins. Otherwise, Picard wins. The payoff is 1 for the winner and -1 for the loser. This is an \emph{asymmetric} game, because the two players have a different number of moves and their payoff functions are different. 

As shown in Section \ref{qip}, both classical and quantum states can be represented in density matrices, while classical and  quantum strategies can be represented as unitary operators. Using the qubit state $|0\rangle$ to represent head-up and $|1\rangle$ to represent tail-up of a penny, the  initial head-up state of the penny is $\rho_0=\begin{bmatrix} 1 & 0 \\ 0 & 0 \end{bmatrix}$; the classical flip unitary operator is $F=\begin{bmatrix} 0 & 1 \\ 1 & 0 \end{bmatrix}$ and the classical no-flip unitary operator is $N=\begin{bmatrix} 1 & 0 \\ 0 & 1 \end{bmatrix}$. 

If both Q and Picard use classical strategies, this three-move game has no \emph{pure classical} NE strategies. However,  it has a continuum of \emph{mixed classical} NE strategies ($\frac{1}{2}$, $\frac{1}{2}$, *)  $\cup$ (*, $\frac{1}{2}$, $\frac{1}{2}$): Q flips the penny with probability $\frac{1}{2}$ in one of his two moves and uses an arbitrary strategy in the other move; Picard flips the penny with probability $\frac{1}{2}$. The result is a tie, where both players have equal probability to win the game. In other words, Q does not have advantage over Picard due to his one extra move.

To understand this continuum of mixed classical NE strategies, we have to explain two observations. First, the state $\rho=\begin{bmatrix} \frac{1}{2} & 0 \\ 0 & \frac{1}{2} \end{bmatrix}$ is a ``stuck state", where no classical or quantum strategy can change the state. 

In the classical case, let an arbitrary mixed classical strategy have probability $p$ to flip the penny and probability $(1-p)$ not to flip the penny (note that pure strategies are a special case where $p=1$ or $p=0$). The strategy transforms $\rho_{i}=\begin{bmatrix} \frac{1}{2} & 0 \\ 0 & \frac{1}{2} \end{bmatrix}$ to 
$\rho_{i+1}=pF\rho_{i} F^{\dagger}+(1-p)N\rho_{i} N^{\dagger}=\begin{bmatrix} \frac{1}{2} & 0 \\ 0 & \frac{1}{2} \end{bmatrix}$.
In the quantum case, let $U=\begin{bmatrix} a & b \\ \bar{b} & -\bar{a} \end{bmatrix}$ be an arbitrary pure quantum strategy, which would transform $\rho_{i}=\begin{bmatrix} \frac{1}{2} & 0 \\ 0 & \frac{1}{2} \end{bmatrix}$ to
$\rho_{i+1}=U\rho_{i} U^{\dagger}=\begin{bmatrix} a & b \\ \bar{b} & -\bar{a} \end{bmatrix}\begin{bmatrix} \frac{1}{2} & 0 \\ 0 & \frac{1}{2} \end{bmatrix}\begin{bmatrix} \bar{a} & b \\ \bar{b} & -a \end{bmatrix}=\begin{bmatrix} \frac{1}{2} & 0 \\ 0 & \frac{1}{2} \end{bmatrix}$.

Second, a mixed classical strategy with $p=\frac{1}{2}$ (half-half) can transform an arbitrary classical state $\rho=\begin{bmatrix} s & 0 \\ 0 & 1-s\end{bmatrix}$ to the stuck state $\begin{bmatrix} \frac{1}{2} & 0 \\ 0 & \frac{1}{2} \end{bmatrix}$ as shown below:

$\frac{1}{2}F\begin{bmatrix} s & 0 \\ 0 & 1-s\end{bmatrix} F^{\dagger}+\frac{1}{2}N\begin{bmatrix} s & 0 \\ 0 & 1-s\end{bmatrix} N^{\dagger}=\begin{bmatrix} \frac{1}{2} & 0 \\ 0 & \frac{1}{2} \end{bmatrix}$. \\


Hence, regardless what classical strategy Q uses in his first move, Picard's half-half mixed classical strategy will transform the penny to the stuck state. Once stuck, Q's second classical strategy can not change the state to a different state. The game ends in a tie where both Picard and Q have equal probability to win the game. Note that one of Q's two moves has to be the half-half mixed strategy. Otherwise, Picard can change his half-half mixed strategy to a different strategy to improve his payoff according to the non-half-half strategy that Q plays. 

The continuum of NE strategies ($\frac{1}{2}$, $\frac{1}{2}$, *)  $\cup$ (*, $\frac{1}{2}$, $\frac{1}{2}$) is the classical ES set for this game. Any classical strategies that is outside this set would give one of the two players  expected payoff that is less than 0. Under selection pressure, this kind of strategies will become extinct in the population.

\section{Quantum Evolutionarily Stable Sets} \label{qess}
Starting with a population of strategies from the mixed classical ES set, we design the following three mutant invasion simulations to investigate the existence of quantum ES set in the quantum penny flip games:
\begin{enumerate}
\item Q's mixed classical strategies in the ES set are invaded by pure quantum strategies to play against Picard's mixed classical strategies in the ES set;
\item Picard's mixed classical strategies in the ES set are invaded by pure quantum strategies to play against Q's mixed classical strategies in the ES set;
\item Q's mixed classical strategies in the ES set are invaded by pure quantum strategies while Picard's mixed classical strategies in the ES set are invaded by mixed two quantum strategies to play against each other.
\end{enumerate}

We designed our EGT model based on the evolutionary framework described by Hofbauer and Sigmund \cite{hofbauer_sigmund} to conduct the designed simulations. In this model, there are two populations $P$ and $K$, one for Picard's and one for Q's strategies. At each generation, every $p\in P$ plays against every $k\in K$.  At each contest, the payoff of $p$ is the probability of the penny's final state ($\rho_{final}$) being measured as $|1\rangle$ minus the probability of it being measured as $|0\rangle$. By contrast, the payoff of $k$ is the probability of $\rho_{final}$ being measured as $|0\rangle$ minus the probability of it being measured as $|1\rangle$. The fitness of $p$ is the average payoff of its contests against all $k\in K$. Similarly, the fitness of $k$ is the average payoff of its contests against all $p\in P$. 

\vspace{-0.3cm}
\small
\begin{align}
f(p)=&\frac{\sum_{k\in K} f_P(p,k)}{|K|}, f_P(p,k)=\rho_{final}(1,1)-\rho_{final}(0,0)\\
f(k)=&\frac{\sum_{p\in P} f_K(p,k)}{|P|}, f_K(p,k)=\rho_{final}(0,0)-\rho_{final}(1,1) 
\label{fitness}
\end{align}
\normalsize
We used the following two-parameter unitary operator to represent a quantum strategy:
\begin{center}
$U(\theta,\phi)=\begin{bmatrix} cos \theta & -e^{i\phi} sin \theta  \\ sin \theta & e^{i\phi} cos \theta \end{bmatrix}.$
\end{center}
where $\theta \in [0,\frac{\pi}{2}]$ and $\phi \in [0,\pi]$. The classical flip strategy $F$ is $U(\frac{\pi}{2},\pi)=\begin{bmatrix} 0 & 1 \\ 1 & 0 \end{bmatrix}$. The classical no-flip strategy $N$ is $U(0,0)=\begin{bmatrix} 1 & 0 \\ 0 & 1 \end{bmatrix}.$

We implemented the EGT model using Holland's Genetic Algorithms (GAs) \cite{holland}. An individual is a linear chromosome that consists of a number of genes. To encode a mixed classical strategy, the gene is the probability ($pro$) to flip the penny. To encode a pure quantum strategy $U$, the genes are the values of $\theta$ and $\phi$. Encoding a mixed-two quantum strategy requires 5 gene values: the probability of applying the first quantum strategy, the $\theta$ and $\phi$ of the first quantum strategy and the $\theta$ and $\phi$ of the second quantum strategy.

The half-half mixed strategy can be coded in two different ways: $p=0.5$ or $U(\frac{\pi}{4}, *)$, where * is a random value between 0 and $\pi$.  This is because when applied to an arbitrary classical state $\rho_i=\begin{bmatrix} s & 0 \\ 0 & 1-s\end{bmatrix}$, $U(\frac{\pi}{4},*)$ 
produces  the same effect as the half-half mixed classical strategy on the diagonal elements: 
\begin{center}
$\rho_{i+1}=U(\frac{\pi}{4},*)\begin{bmatrix} s & 0 \\ 0 & 1-s \end{bmatrix}U(\frac{\pi}{4},*)^{\dagger}=\begin{bmatrix} \frac{1}{2} & s-\frac{1}{2} \\   s-\frac{1}{2} & \frac{1}{2} \end{bmatrix}$.
\end{center}
When measured, the penny collapses to $|0\rangle$ and $|1\rangle$ with equal probability.

Note that $\rho_{i+1}$ is not the same as the ``stuck state", since its off-diagonal values are not zero. If $\rho_{i+1}$ is an intermediate state of the penny, quantum strategies can change the state although classical strategies cannot. Hence, the final measurement of the penny may not be a tie. However, in simulations 1 \& 2, one of the two players remains using classical strategies, which can not modify $\rho_{i+1}$. The U($\frac{\pi}{4}$,*) strategy, therefore, behaves identical to the classical half-half mixed strategy.

The system uses the following standard GA operators:
\begin{itemize}
\item Binary tournament selection: two chromosomes are randomly selected from a population and the one with higher fitness is the winner.
\item Gaussian mutation: each gene value of a selected chromosome has a specified probability to be mutated by adding random values from a Gaussian distribution under a specified standard deviation to produce a new offspring.
\item Average crossover: the genes of two selected chromosomes are averaged to produce one offspring.
\end{itemize}

\begin{figure}[!htp]
\vspace{-0.5cm}
\centering
\includegraphics[height=4in, width=3.5in]{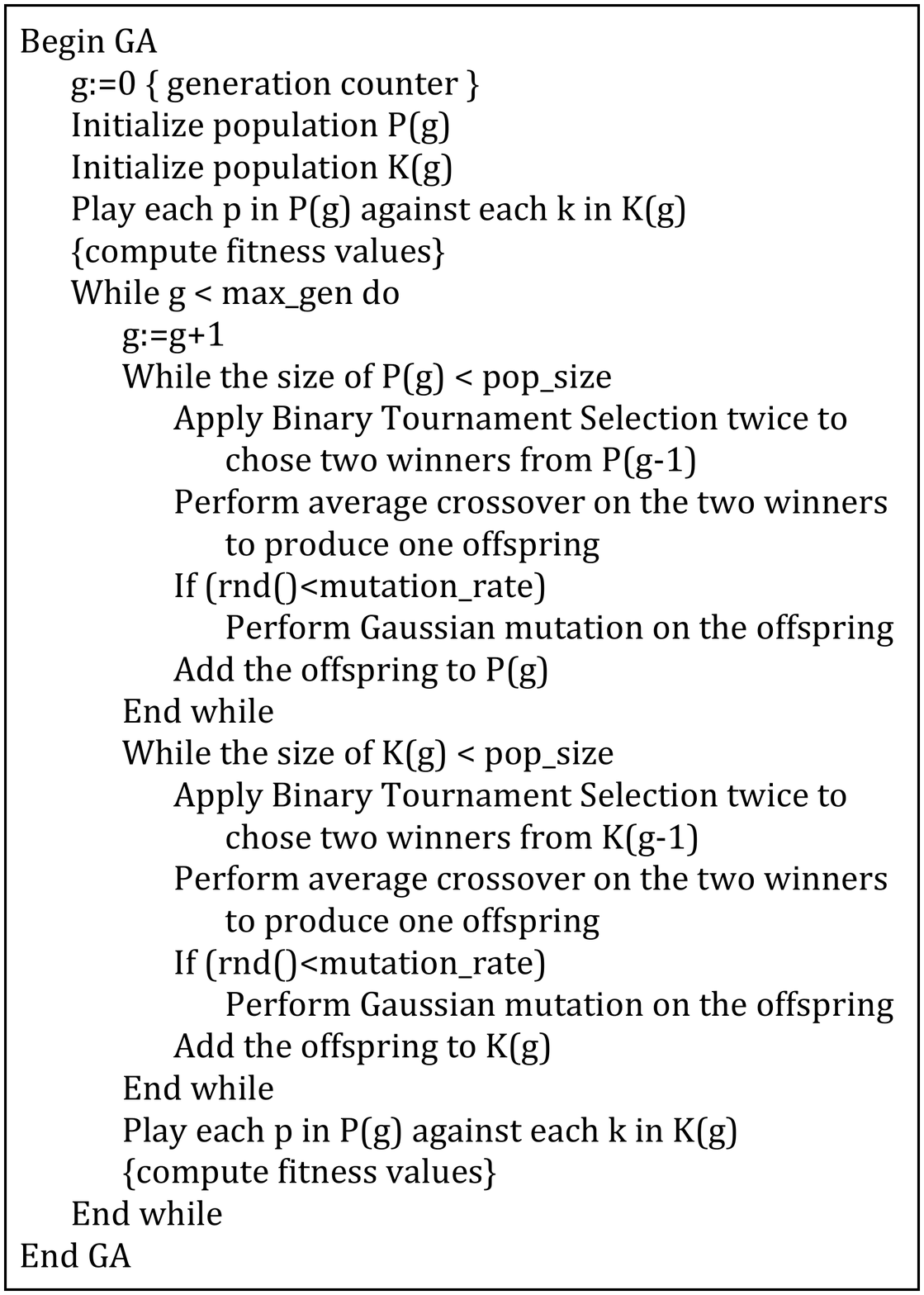}
\vspace{-0.2cm}
\begin{center}
\caption{ The genetic algorithm system work flow.}
\end{center}
\vspace{-0.2cm}
\label{ga}
\end{figure}

Figure \ref{ga} gives the GA system workflow. Initially, two populations, P and K, are seeded with classical strategies from the ES set. The two populations of strategies then play against each other and their fitness values are evaluated according to equations 14 \&  \ref{fitness}. Based on the fitness, fitter strategies are selected to perform average crossover and Gaussian mutation to produce offspring for the next generation. This process of selection-reproduction-evaluation is repeated for many generations until the maximum number of generation is reached.

Table \ref{para} lists the GA parameters used to run the simulations. The maximum number of generation is 500 for simulations 1 \& 2 and 10,000 for simulation 3. We present simulation 1 \& 2 and their results in the following subsection. In Section \ref{mixed_quantum}, we present and analyze the results of simulation 3.

\begin{table}[!htb]
\centering
\caption{GA Parameter values to run the simulations.}
\begin{tabular}{|c|c|c|c|} \hline
{\bf parameter}&{\bf value} & {\bf parameter}&{\bf value}\\ \hline\hline
pop\_size&50& max\_gen&  500/10,000 \\\hline
mutation\_rate& 20\% & Gaussian mutation std & 0.2 \\\hline
\end{tabular}
\label{para}
\end{table}


\subsection{Quantum Strategies Invade One of the Two Players' Classical Strategies}\label{pure_quantum}

As analyzed in Section \ref{pqgame}, the game has a mixed classical ES set $(\frac{1}{2}, \frac{1}{2},*) \cup (*,\frac{1}{2},\frac{1}{2})$. We encode the half-half strategy as $U(\frac{\pi}{4}, *)$  in the population that is to be invaded by quantum strategies, which can be population P or K, hence any $U(\theta\ne \frac{\pi}{4}, *)$ is a quantum mutant strategy. By contrast, the half-half strategy in the population that is not invaded by quantum strategies, which can be population P or K, is coded as $pro=\frac{1}{2}$. In this way, the $pro$ can be mutated to any probability to adapt to the invasion that is taking place in the other population (see the work flow in Figure \ref{ga}).  
 
\subsubsection{Simulation 1:}
\begin{figure}[!htp]
\begin{minipage}[t]{0.49\linewidth}
\centerline{\includegraphics[width=6cm,height=4.5cm]{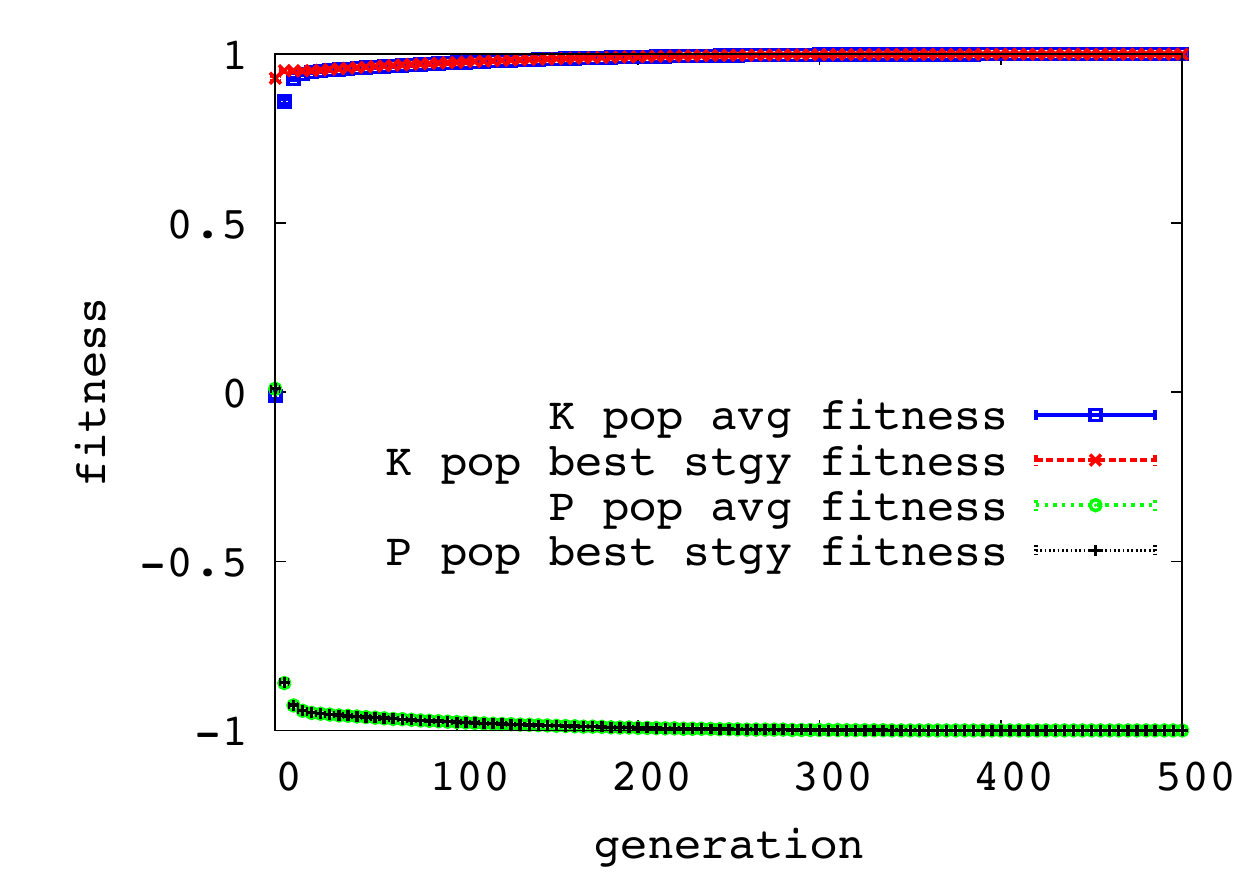}}
\caption{P and K avg population fitnesses.
\label{case1.1}}
\end{minipage}
\begin{minipage}[t]{0.49\linewidth}
\centerline{\includegraphics[width=7.0cm,height=4.5cm]{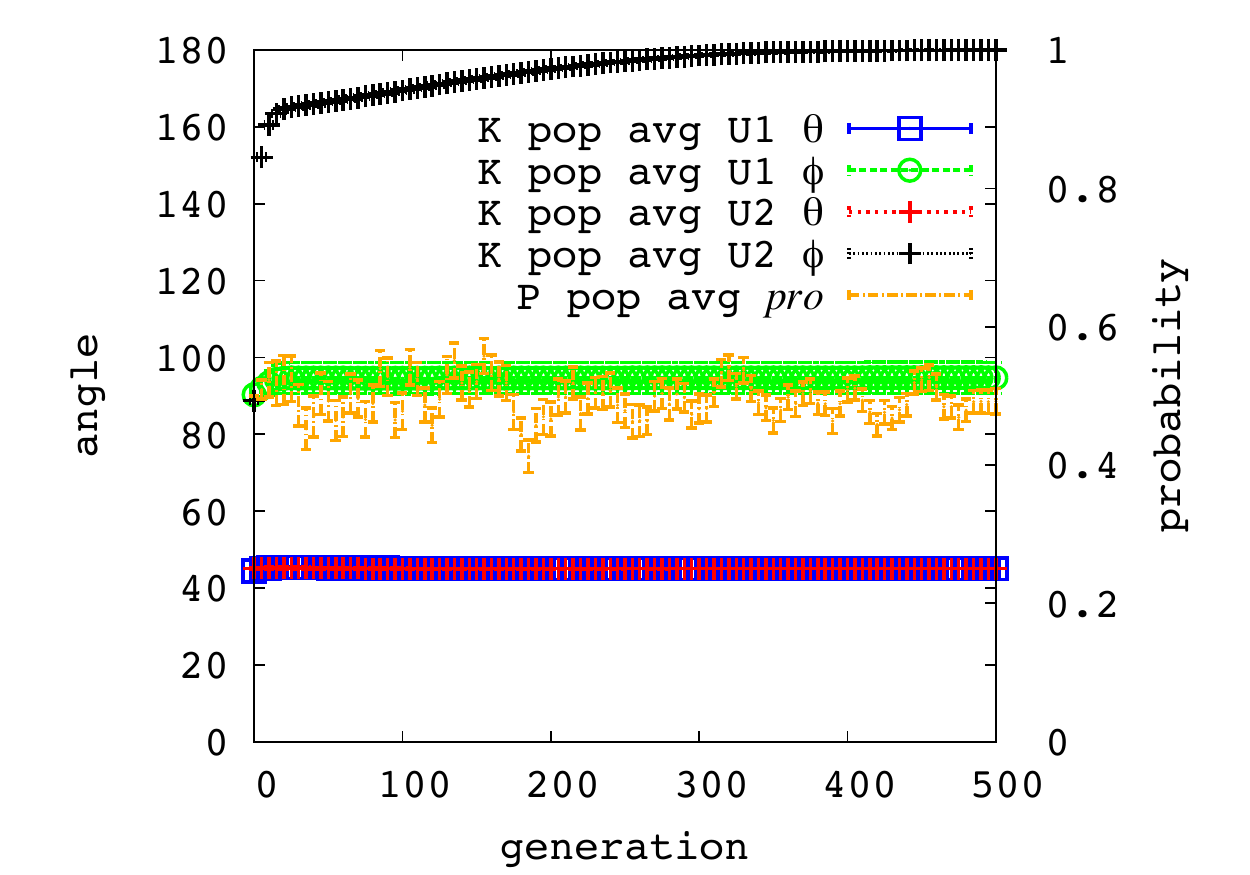}}
\caption{P and K strategies contents.
\label{case1.2}}
\end{minipage}
\end{figure}

In this simulation, $K$  is invaded. The initial $K$ population is therefore seeded with each $k \in K$ having two unitary operators $U1$ and $U2$, one for each of its two moves, as $U(\frac{\pi}{4}, *)$ and $U(*,*)$, where * is a value randomly chosen such that $0 \le \theta \le \frac{\pi}{2}$ and $0 \le \phi \le \pi$. By contrast, all $p\in P$ are identical with $pro=0.5$. We made 100 simulation runs and the population average fitness with the standard error of the mean (SEM) are given in Figure \ref{case1.1}. 

At generation 0, both populations were seeded with mixed classical strategies from the ES set, hence all individuals received payoff  0. Once some of the $U$ in the K population and some of the $pro$ in the P population were mutated, the average fitness of the K population increased very quickly, while the average fitness of the P population dropped very quickly. At generation 400, the average fitness of both populations converged, where all quantum strategies in the K population received fitness close to 1 (0.9999) while all mixed classical strategies in the P population received fitness close to -1 (-0.9999). 


Figure \ref{case1.2} shows the evolved strategies in both populations. For the K population, they are the average $\theta$ and $\phi$ of $U1$ and $U2$ with the SEM. For the P population, they are the average $pro$ with the SEM.

At generation 0, the K population consisted of two types of individuals: $[U(\frac{\pi}{4}, *),U(*,*)]$ and $[U(*,*),U(\frac{\pi}{4}, *)]$, where * is a randomly generated value that satisfies the constraints, $0 \le  \theta \le \frac{\pi}{2}$ and $0 \le \phi \le \pi $. All $pro$ in the P population, on the other hand, are 0.5. 

After the evolution started, the average $\theta$ of the two quantum strategies $U1$ and $U2$ in the K population remained as $\frac{\pi}{4}$ while the average $\phi$ of $U1$ stayed around $\frac{\pi}{2}$ but with a large standard error. By contrast, the average $\phi$ of $U2$ grew and converged to $\pi$ around generation 400. Meanwhile, the average $pro$ of the P population fluctuated between 0.4 and 0.6, but did not converge to a particular value.
We analyze the penny states using the evolved strategy [$U(\frac{\pi}{4},*), pro, U(\frac{\pi}{4},\pi)$]. 
Given the head-up state, $\rho_0=\begin{bmatrix} 1 & 0 \\ 0 & 0 \end{bmatrix}$, the state of the penny after Q applied $U(\frac{\pi}{4},*)$ is 
$\rho_1=U(\frac{\pi}{4},*)\rho_0U(\frac{\pi}{4},*)^{\dagger}=\begin{bmatrix} \frac{1}{2} & \frac{1}{2} \\ \frac{1}{2} & \frac{1}{2} \end{bmatrix}$. Next, Picard applied his mixed classical strategy  ($pro$) to transform the penny state to $\rho_2=proF\rho_1 F^{\dagger}+(1-pro)N\rho_1N^{\dagger}=\begin{bmatrix} \frac{1}{2} & \frac{1}{2} \\ \frac{1}{2}& \frac{1}{2} \end{bmatrix}$, which is identical to $\rho_1$. In other words, Picard's mixed classical strategy could not change the state of the penny. Since all $pro$ values produce the same $\rho_2$, the $pro$ in P population had a wide range between 0 and 1 with average around 0.4 and 0.6, as that shown in Figure \ref{case1.2}. Note that unlike the ``stuck state'', the phase (non-diagonal value) of $\rho_2$ is not zero. Although classical strategies cannot change the state, quantum strategies can modify the phase and through the effect of interference (see Section \ref{qip}), the state of the penny may change.

If both players had only one move, the game would have ended and both players received the same payoff of 0. 
However, this is not the case, as Q had one more move. He then applied $U(\frac{\pi}{4},\pi)$, whose interference effect changed the state to: $\rho_3=U(\frac{\pi}{4},\pi)\rho_2U(\frac{\pi}{4},\pi)^{\dagger}=\begin{bmatrix} 1& 0 \\ 0 & 0 \end{bmatrix}$. When measured, the penny collapses to $|0\rangle$ with probability 1. The game therefore ended with Q receiving payoff 1 and Picard receiving payoff -1.  While the extra move did not give Q advantage in the classical version of the game, it helped him to win in this version of quantum penny flip game with certainty.


Why did the $\phi$ of $U1$ have such a large standard error? Given the initial head-up state $\rho_0=\begin{bmatrix} 1 & 0 \\ 0 & 0 \end{bmatrix}$, the state of penny after applying an arbitrary  strategy $U(\theta,\phi)=\begin{bmatrix} cos \theta & -e^{i\phi} sin \theta  \\ sin \theta & e^{i\phi} cos \theta \end{bmatrix}$ is:
$\rho_{1}=U(\theta,\phi)\rho_{0} U(\theta,\phi)^{\dagger}=\begin{bmatrix} cos \theta & -e^{i\phi} sin \theta  \\ sin \theta & e^{i\phi} cos \theta \end{bmatrix}$ $\begin{bmatrix} 1& 0 \\ 0 & 1 \end{bmatrix}\begin{bmatrix} cos \theta & sin \theta  \\  -e^{-i\phi} sin \theta & e^{-i\phi} cos \theta \end{bmatrix}=\begin{bmatrix} cos \theta ^2 & sin\theta cos\theta \\ sin\theta cos\theta & sin\theta^2 \end{bmatrix}$. In other words, $\phi$ of $U1$ has no impact on $\rho_1$. Since all U($\frac{\pi}{4}$,*) would produce $\rho_1=\begin{bmatrix} \frac{1}{2} & \frac{1}{2} \\ \frac{1}{2} & \frac{1}{2} \end{bmatrix}$, the $\phi$ can be any value between 0 and $\pi$. Consequently, they average to $\frac{\pi}{2}$ with a large standard error as that shown in Figure \ref{case1.2}.

The set of strategy pairs $[U(\frac{\pi}{4}, *), U(\frac{\pi}{4}, \pi)]$ are the winning quantum strategies for Q, regardless what mixed classical strategies Picard uses. Moreover, any $[U(\theta\ne \frac{\pi}{4}, *), U(\theta\ne \frac{\pi}{4}, \phi\ne \pi)]$ would give Q a lower payoff. The continuum of $[U(\frac{\pi}{4}, *), *, U(\frac{\pi}{4}, \pi)]$ 
is therefore the ES set for this version of the quantum penny flip game. 

\subsubsection{Simulation 2:}
\begin{figure}[!htp]
\begin{minipage}[t]{0.49\linewidth}
\centerline{\includegraphics[width=6.0cm,height=4.5cm]{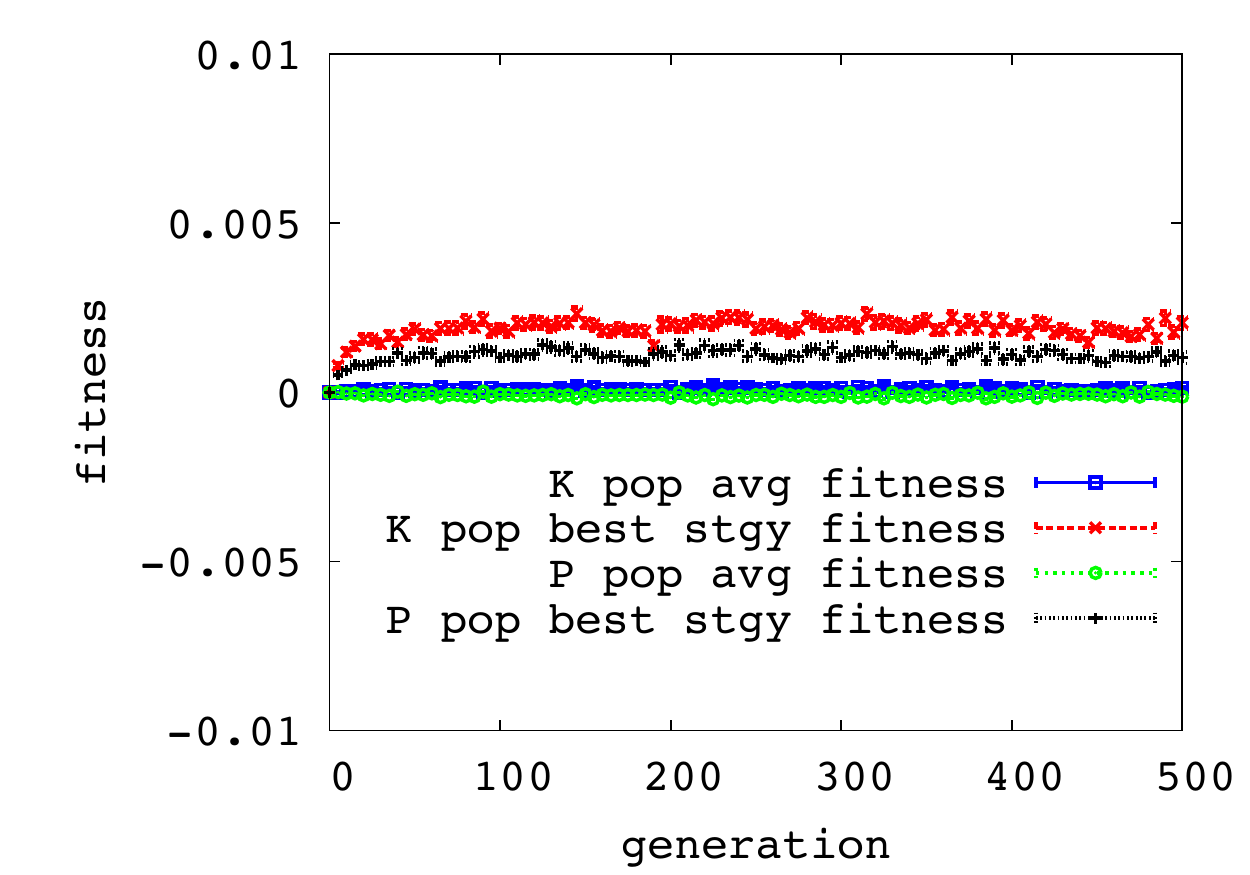}}
\caption{P and K avg population fitnesses.
\label{case2.1}}
\end{minipage}
\begin{minipage}[t]{0.49\linewidth}
\centerline{\includegraphics[width=7cm,height=4.5cm]{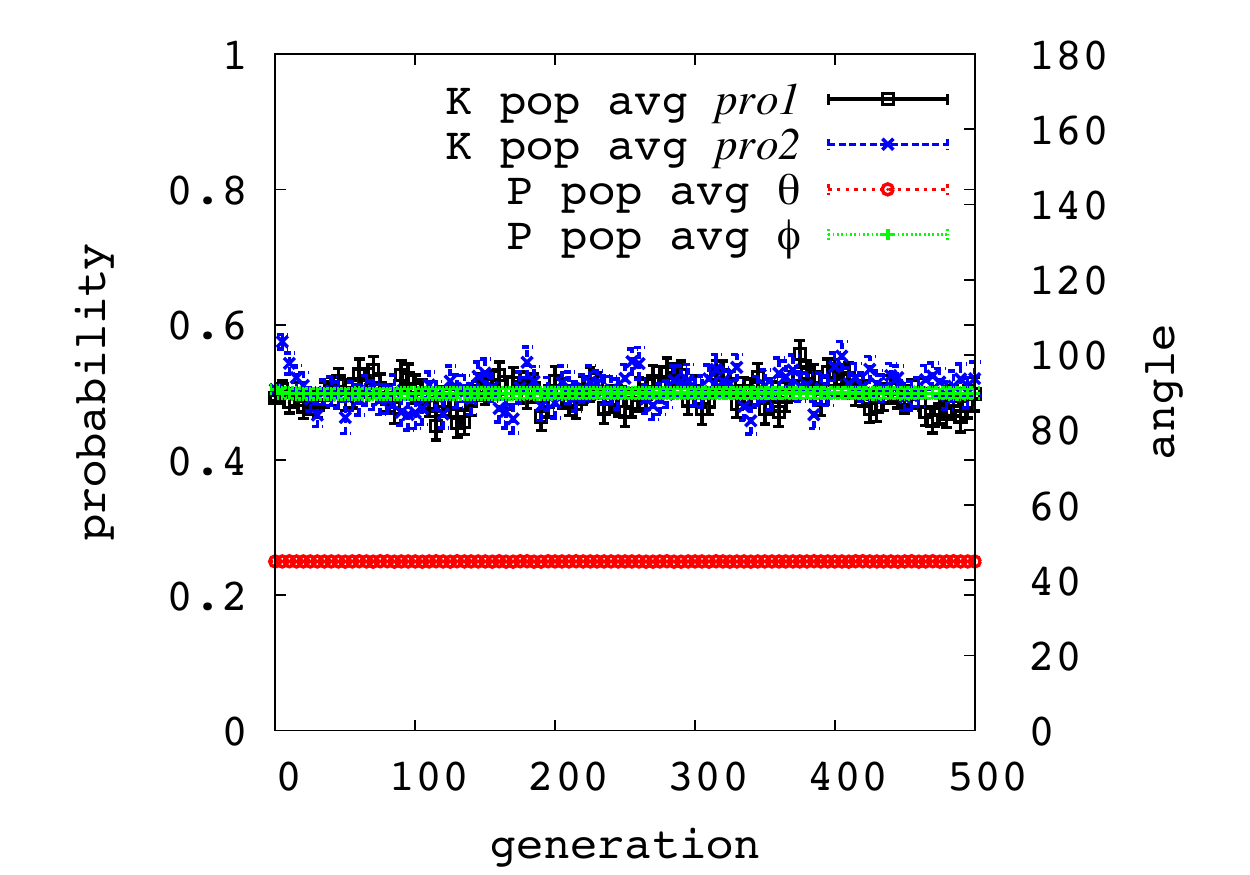}}
\caption{P and K strategies contents.
\label{case2.2}}
\end{minipage}
\end{figure}

In this simulation, $P$  is invaded. The initial $P$ population is therefore seeded with each $p$ as the unitary operator $U(\frac{\pi}{4}, *)$, where * is a random value between 0 and $\pi$. By contrast, the K population is seeded with each $k$ as either $[pro_1=0.5,pro_2=*]$ or $[pro_1=*,pro_2=0.5]$, where * is a random value between 0 and 1. We made 100 simulation runs and the results are presented in Figure \ref{case2.1} and Figure \ref{case2.2}.  

Figure \ref{case2.1} shows that during the entire simulation of 500 generations, the average fitness for both populations stayed close to 0. 
Figure \ref{case2.2} shows that the dominating strategy in the P population converged to $U(\frac{\pi}{4}, \frac{\pi}{2})$ while the average $pro_1$ and $pro_2$ in the K population fluctuated between 0.4 and 0.6 but did not converge to a particular value.


To analyze the ES set for this version of quantum penny flip game, we evaluate the penny states under the three evolved operations [$pro_1$ ,U($\frac{\pi}{4}$, $\frac{\pi}{2}$), $pro_2$].  With initial state $\rho_0=\begin{bmatrix} 1 & 0 \\ 0 & 0 \end{bmatrix}$, after Q applied his mixed classical strategy $pro_1$, the state of the penny is $\rho_1=pro_1F\rho_0 F^{\dagger}+(1-pro_1)N\rho_0N^{\dagger}=\begin{bmatrix} pro_1 & 0 \\ 0& 1-pro_1 \end{bmatrix}$. Next, Picard's $U(\frac{\pi}{4}, \frac{\pi}{2})$ would transform the state to 
$\rho_2=U(\frac{\pi}{4}, 0)\rho_1U(\frac{\pi}{4}, 0)^{\dagger}=\frac{1}{2} \begin{bmatrix} 1 & 2pro_{1}-1 \\   2pro_{1}-1 & 1 \end{bmatrix}$. In fact, Picard can use any $U(\frac{\pi}{4},*)$ to transform the penny to the same state, since $\phi$ has no impact on the state transformation.  We have examined the P population and found that the $\phi$ values spread between 0 and $\pi$, hence averaged to $\frac{\pi}{2}$. With all 100 simulation runs having average $\phi \approx \frac{\pi}{2}$, their average is also close to  $\frac{\pi}{2}$ with a small standard error, as that shown in Figure \ref{case2.2}.

Finally, Q applied his second mixed classical strategy $pro_2$ to transform the penny to $\rho_3=pro_2F\rho_2 F^{\dagger}+(1-pro_2)N\rho_2N^{\dagger}=\frac{1}{2}\begin{bmatrix} 1 & 2pro_{1}-1 \\  2pro_{1}-1 & 1 \end{bmatrix}$, which is identical to $\rho_2$. Once measured, the penny collapsed to $|0\rangle$ and $|1\rangle$ with equal probability, hence both players received expected payoff 0. 
Since neither $pro_1$ nor $pro_2$ has impact on final state, hence the measurement result, their values can have a wide range between 0 and 1. Consequently, their averages are around 0.4 and 0.6, as that shown in Figure \ref{case2.2}. 

This result is similar to the classical version of the game in that no matter what mixed classical strategies Q used in his first move, Picard's $U(\frac{\pi}{4},*)$ half-half strategy will transform the penny to a stuck state, where Q's second classical strategy could not change the state to other state. 
The quantum version of the game therefore has the same ES set as that of the classical version of the game: $[\frac{1}{2},U(\frac{\pi}{4},*), *] \cup [*,U(\frac{\pi}{4}, *), \frac{1}{2}]$. 

Classical strategies are a proper subset of quantum strategies. The ES set in the classical version of a game therefore does not need to be the ES set in the quantum version of the game.  This is the case in simulation 1 where the mixed classical strategies in the ES set are replaced by quantum strategies in that version of quantum game. However, in simulation 2, the  mixed classical ES set remain as the ES set in this version of quantum game.

\subsection{Quantum Strategies Invade Both Players' Classical Strategies}\label{mixed_quantum}
In this simulation, the mixed classical strategies in the $K$ population are invaded by pure quantum strategies, while the mixed classical strategies in the $P$ population are invaded by mixed-two quantum strategies. We designed this simulation as an extension of simulation 1 to explore whether mixed-two quantum strategies can provide better payoff than mixed-two classical strategies for Picard  when playing against Q's pure quantum strategies. 

\begin{figure}[!htp]
\centering
\includegraphics[height=2.0in, width=3.2in]{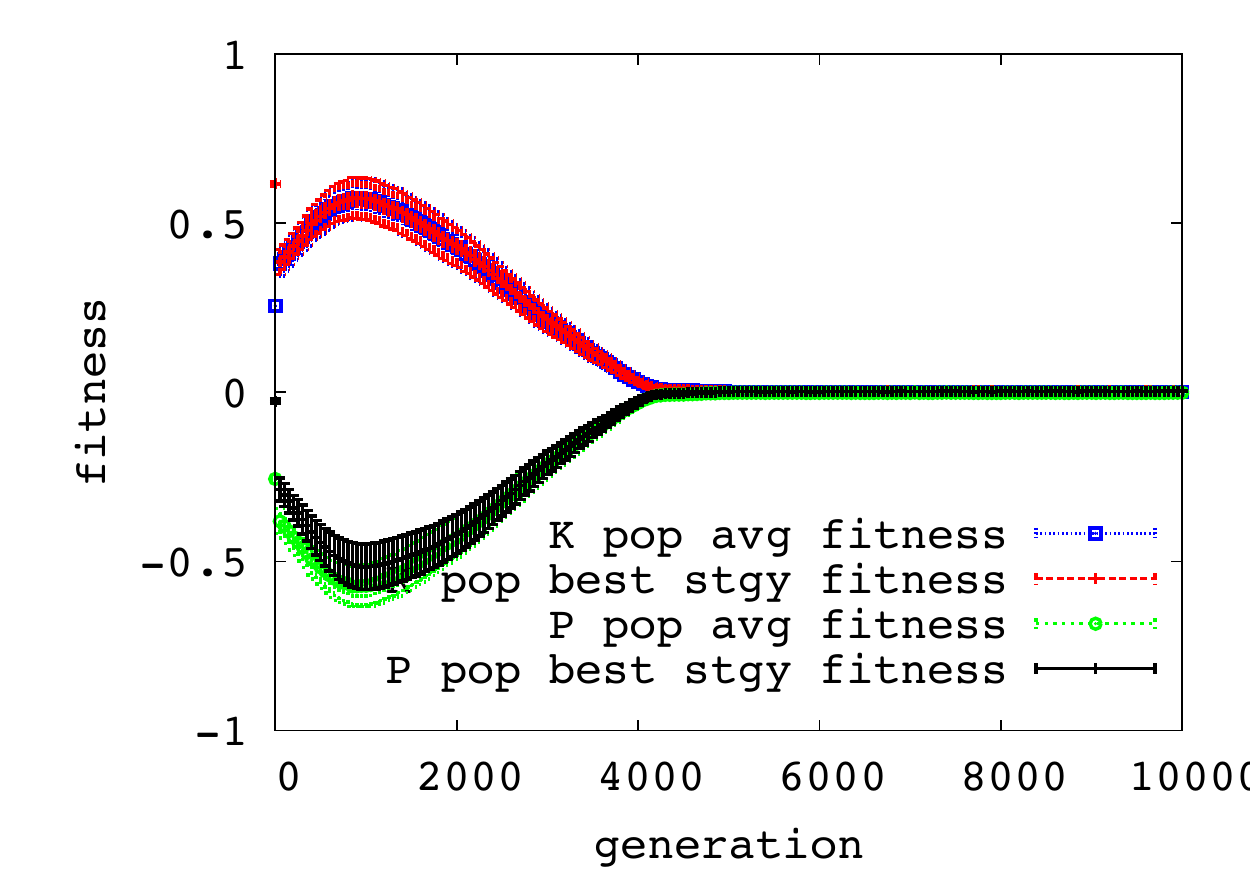}
\vspace{-0.2cm}
\caption{P and K population average fitness.}
\vspace{-0.2cm}
\label{case3.1}
\end{figure}

Similar to simulation 1, the initial $K$ population is seeded with two types of individuals: [U($\frac{\pi}{4}$,*), U(*,*)] and [U(*,*), U($\frac{\pi}{4}$, *)]. 
Unlike simulation 1, all $p\in P$ in this simulation are $[pro=0.5, U(\frac{\pi}{2}, \pi),U(0,0)]$, where $U(\frac{\pi}{2},\pi)=\begin{bmatrix} 0 & 1 \\ 1 & 0 \end{bmatrix}$ is the classical flip strategy and $U(0,0)=\begin{bmatrix} 1 & 0 \\ 0 & 1 \end{bmatrix}$ is  the classical no-flip strategy. In this way, 
the quantum strategies in both populations can evolve against each other to improve their payoffs. 
We made 100 simulation runs and the average population fitness with the SEM are presented in Figure \ref{case3.1}. 

Similar to the two previous simulations, both populations had the same initial average fitness of  0. However, once evolution started, the average fitness of K population increased while the average fitness of P population decreased. This trend continued until generation 800 when the trend is reversed: the average fitness of P population grew while the average fitness of K population shrank. Around generation 4,000, the average fitness of both populations converged to 0, indicating all strategies in both populations give equal probability to win the game. 

Compared to the two previous simulations, the populations in this simulation took 4 times longer to converge. This is because the number of strategy parameters in this simulation (Q has 4 parameters while Picard has 5 parameters) is larger than that in the two previous simulations (Q has 4 or 2 parameters while Picard has 1 or 2 parameters). With a larger parameters space, it took evolution longer to find the stable strategies for this version of quantum penny flip game.

To understand the dynamics of strategies changes during the game, we examined the 100 evolved quantum strategies in both populations. We found that they can be grouped into 4 categories as shown in Table \ref{cases}.  The descriptions of these quantum strategies are given in Table \ref{stgy}. Note that  $\sigma_1, \sigma_2, \sigma_3$ are Pauli matrices. We represent $\sigma_2$ as $\begin{bmatrix} 0 & -1 \\ 1 & 0 \end{bmatrix}$, which behaves identical to $\begin{bmatrix} 0 & -i \\ i & 0 \end{bmatrix}$ when applied to an arbitrary state $\rho=\begin{bmatrix} s & a \\  \bar{a} & 1-s \end{bmatrix}$: $\begin{bmatrix} 0 & -1 \\  1 & 0 \end{bmatrix}\begin{bmatrix} s & a \\  \bar{a} & 1-s \end{bmatrix}\begin{bmatrix} 0 & -1 \\  1 & 0 \end{bmatrix}^{\dagger}=\begin{bmatrix} 0 & -i \\  i & 0 \end{bmatrix}\begin{bmatrix} s & a \\  \bar{a} & 1-s \end{bmatrix}\begin{bmatrix} 0 & -i \\  i & 0 \end{bmatrix}^{\dagger}=\begin{bmatrix} 1-s & -\bar{a} \\  -a & s \end{bmatrix}$.

\begin{table}[!ht]
\caption{ Four categories of the evolved quantum strategies.}
\centering
\begin{tabular}{|c|c|c|c|}
\hline
category & Q's pure quantum strategies &  Picard's mixed quantum strategies & qty\\
\hline\hline
1&U1=U($\frac{\pi}{4}$,*), U2=U(*,$\frac{\pi}{2}$) &  mixed $\sigma_1$ and $\sigma_3$ & 43\\
\hline
2&U1=U($\frac{\pi}{4}$,*), U2=U(*,$\frac{\pi}{2}$)&  mixed $\sigma_2$ and  $I$ & 53\\
\hline
3&U1=U(0,*), U2=H &  mixed $\sigma_3$ and $\sigma_2$ & 3\\
\hline
4&U1=U($\frac{\pi}{2}$,*), U2=U($\frac{\pi}{4}$,0) & mixed $I$ and $\sigma_1$& 1 \\
\hline
\end{tabular}
\label{cases}
\end{table}

\begin{table}[!ht]
\caption{Descriptions of the evolved quantum strategies.}
\centering
\begin{tabular}{|c|c|}
\hline
strategy & unitary matrix\\
\hline\hline
$\sigma_1$ & $U(\frac{\pi}{2},\pi)$ =  $\begin{bmatrix} 0 & 1 \\ 1 & 0 \end{bmatrix}$ \\
\hline
$\sigma_2$ & $U(\frac{\pi}{2},0)$ =   $\begin{bmatrix} 0 & -1 \\ 1 & 0 \end{bmatrix}$\\
\hline
$\sigma_3$ & $U(0,\pi)$ = $\begin{bmatrix} 1 & 0 \\ 0 & -1 \end{bmatrix}$\\
\hline
$I$ & $U(0,0)$ =  $\begin{bmatrix} 1 & 0 \\ 0 & 1 \end{bmatrix}$\\
\hline
$Hadamard(H)$ &  $U(\frac{\pi}{4},\pi)$ =  $\frac{1}{\sqrt{2}}\begin{bmatrix} 1 & 1 \\ 1 & -1 \end{bmatrix}$\\
\hline
\end{tabular}
\label{stgy}
\end{table}

\newpage

For each of the 4 categories, we analyze the evolved quantum strategies in the following subsections.

\subsubsection{Category 1 Quantum Strategies:} 
\begin{figure}[!htb]
\begin{minipage}[t]{0.49\linewidth}
\centerline{\includegraphics[width=6.5cm,height=4.0cm]{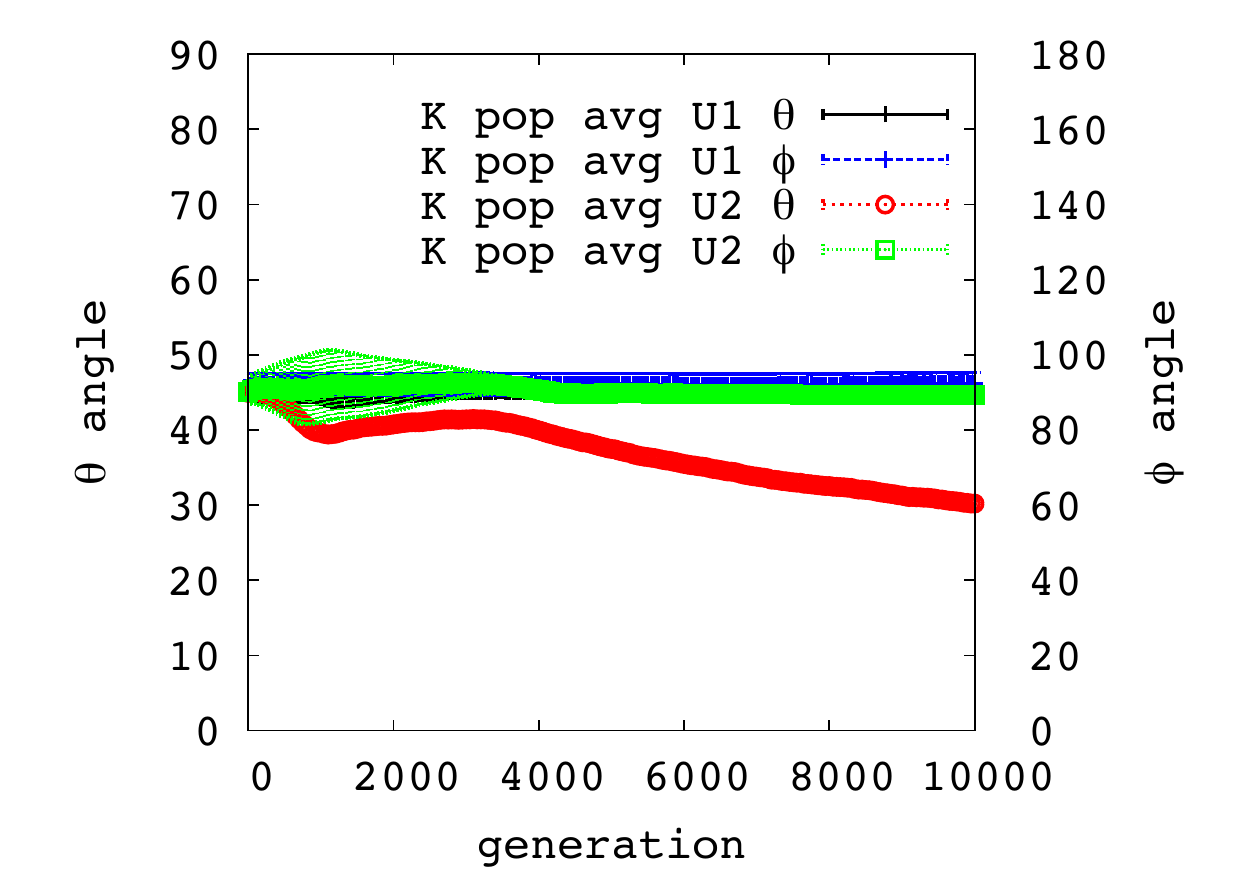}}
\caption{Evolved Q strategies
\label{case1_q}}
\end{minipage}
\begin{minipage}[t]{0.49\linewidth}
\centerline{\includegraphics[width=6.5cm,height=4.0cm]{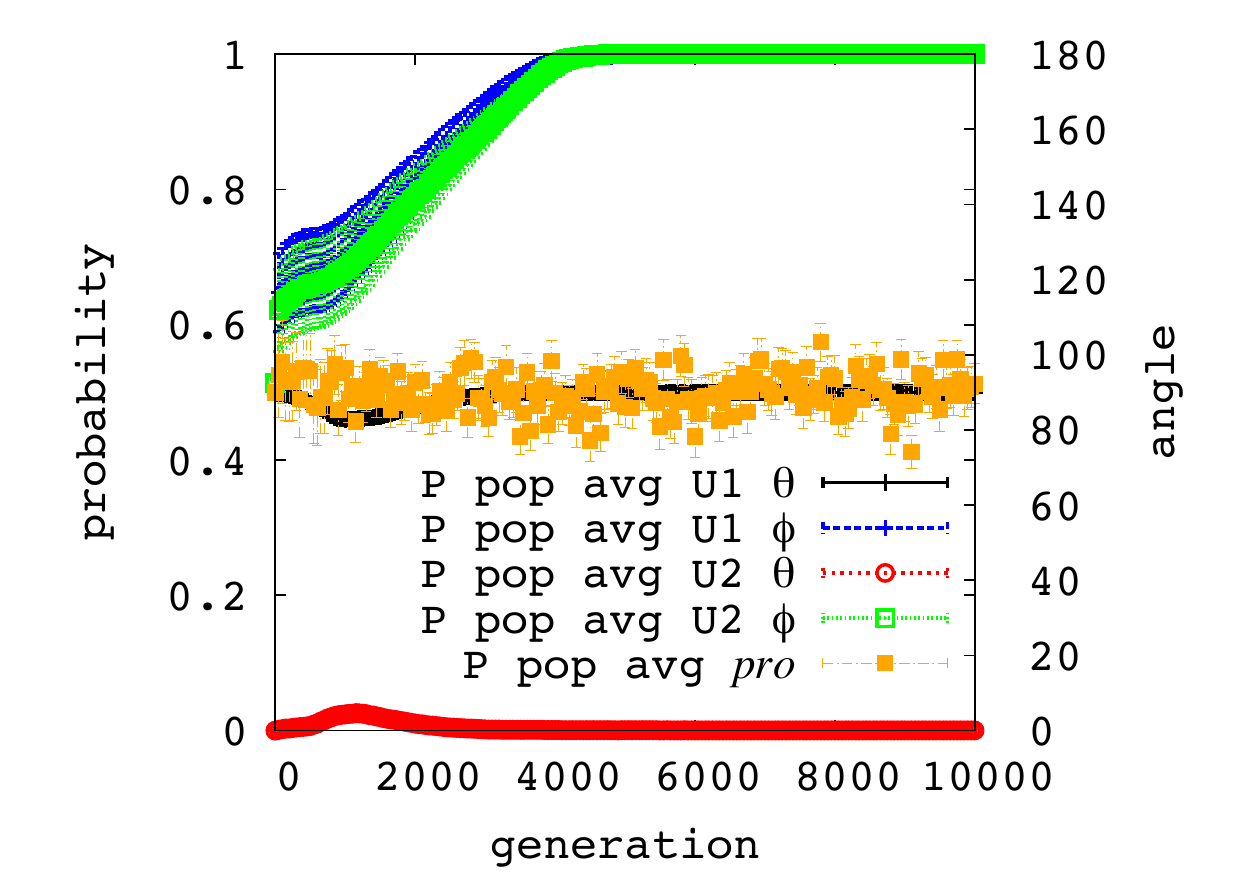}}
\caption{Evolved Picard strategies.
\label{case1_p}}
\end{minipage}
\end{figure}

Among the 100 simulation runs, 43 of them converged to strategies in category 1, where Q used U($\frac{\pi}{4}$,*) and  U(*, $\frac{\pi}{2}$)  to play against Picard's mixed $\sigma_1$ and $\sigma_3$ strategies. As analyzed in simulation 1, $\phi$ of $U1$ does not have impact on $\rho_1$. Q could therefore apply any U($\frac{\pi}{4}$,*) to transform the initial state $\rho_0=\begin{bmatrix} 1 & 0 \\ 0 & 0 \end{bmatrix}$ to $\rho_1=U(\frac{\pi}{4},*)\begin{bmatrix} 1 & 0 \\ 0 & 0 \end{bmatrix}U(\frac{\pi}{4},*)^{\dagger}=\frac{1}{2}\begin{bmatrix} 1 & 1 \\ 1 & 1 \end{bmatrix}$. 
Next, Picard applied mixed $\sigma_1$ and $\sigma_3$ to transform the penny to $\rho_2=(pro) \sigma_1\rho_1 \sigma_1^{\dagger}+(1-pro)\sigma_3\rho_1\sigma_3^{\dagger}=\frac{1}{2}\begin{bmatrix} 1 & 2pro-1 \\  2pro-1 & 1 \end{bmatrix}$. Finally, Q applied U(*,$\frac{\pi}{2}$) and transformed the penny to $\rho_3= U(*,\frac{\pi}{2}) \rho_2 U(*,\frac{\pi}{2})^{\dagger} =\frac{1}{2}\begin{bmatrix} 1 & a \\ \bar{a} & 1 \end{bmatrix}$. 
When measured, the penny collapses to $|0\rangle$ and to $|1\rangle$ with equal probability, hence both players received expected payoff 0.

Note that $\theta$ of Q's $U2$ has no impact on the final state $\rho_3$, hence the measurement of the penny.  This is because given an arbitrary strategy $U(\theta,\phi)=\begin{bmatrix} cos \theta & -e^{i\phi} sin \theta  \\ sin \theta & e^{i\phi} cos \theta \end{bmatrix}$, the state $\rho_{3}=U(\theta,\phi)\rho_{2} U(\theta,\phi)^{\dagger}=\begin{bmatrix} cos \theta & -e^{i\phi} sin \theta  \\ sin \theta & e^{i\phi} cos \theta \end{bmatrix}\frac{1}{2} \\
\begin{bmatrix} 1& 2pro-1 \\  2pro-1 & 1 \end{bmatrix}\begin{bmatrix} cos \theta & sin \theta  \\  -e^{-i\phi} sin \theta & e^{-i\phi} cos \theta \end{bmatrix} = \frac{1}{2} \begin{bmatrix} sin\theta^2+cos\theta^2 & a \\ \bar{a} & sin\theta^2+cos\theta^2 \end{bmatrix} \\ = \frac{1}{2}\begin{bmatrix} 1 & a \\ \bar{a} & 1 \end{bmatrix}$. Since all U(*,$\frac{\pi}{2}$) produce the $\rho_3$ that give the same measurement result, $\theta$ of $U2$ can be any value between 0 and $\frac{\pi}{2}$. In our simulation runs, $\theta$ of $U2$ averaged to $\frac{\pi}{3}$ (see Figure \ref{case1_q}). Similarly, Picard's $pro$ also does not have impact on $\rho_3$. Figure \ref{case1_p} shows that Its values frustrated between 0 and 1 and averaged to between 0.4 and 0.6.

\subsubsection{Category 2 Quantum Strategies:}
\begin{figure}[!htb]
\begin{minipage}[t]{0.49\linewidth}
\centerline{\includegraphics[width=6.5cm,height=4.0cm]{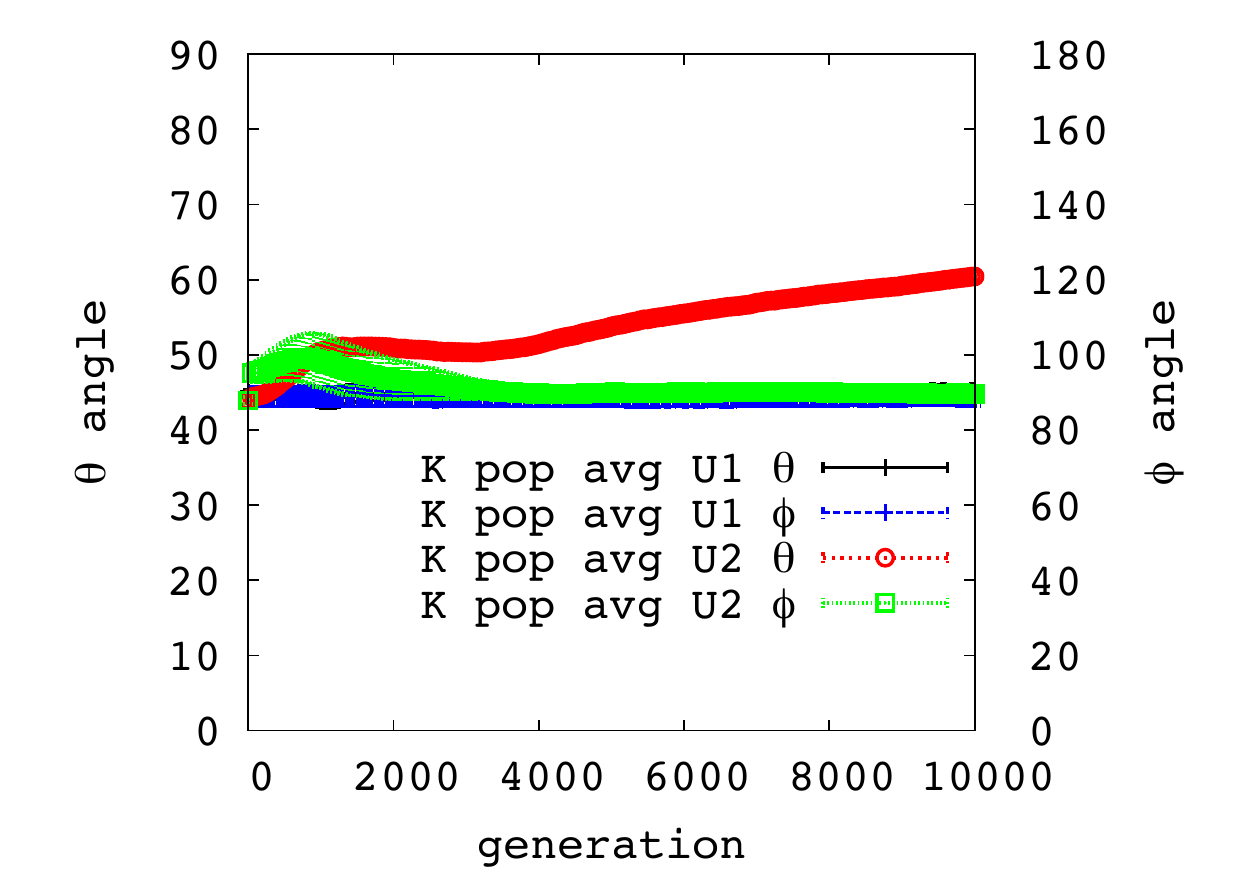}}
\caption{Evolved Q strategies
\label{case2_q}}
\end{minipage}
\begin{minipage}[t]{0.49\linewidth}
\centerline{\includegraphics[width=6.5cm,height=4.0cm]{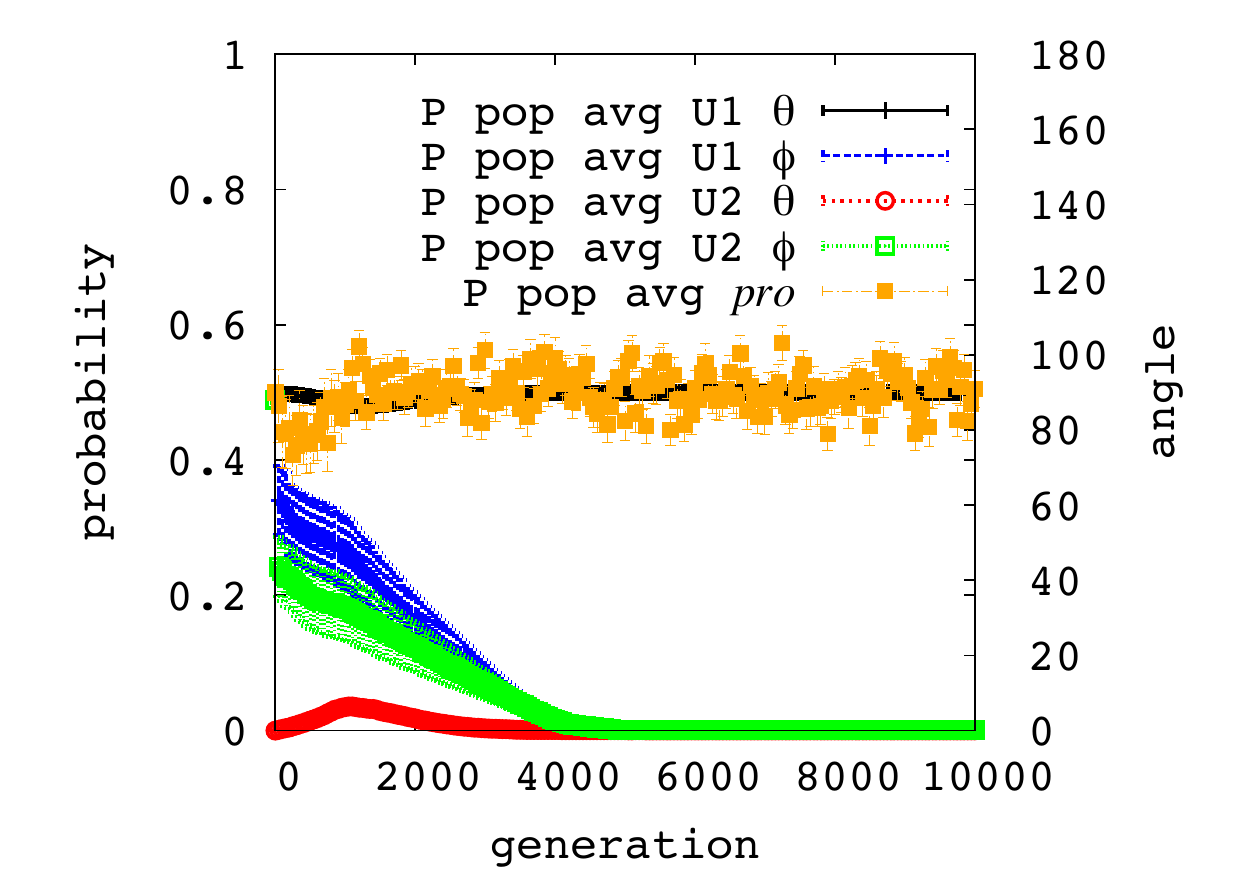}}
\caption{Evolved Picard strategies.
\label{case2_p}}
\end{minipage}
\end{figure}

53 of the 100 simulation runs converged to strategies in category 2, where Q used U($\frac{\pi}{4}$,*) and U(*,$\frac{\pi}{2}$) to play against Picard's mixed $\sigma_2$ and $I$. With the initial state $\rho_0=\begin{bmatrix} 1 & 0 \\ 0 & 0 \end{bmatrix}$, after Q applied U($\frac{\pi}{4}$, *), the state of the penny was $\rho_1=U(\frac{\pi}{4},*)\begin{bmatrix} 1 & 0 \\ 0 & 0 \end{bmatrix}U(\frac{\pi}{4},*)^{\dagger}=\frac{1}{2}\begin{bmatrix} 1 & 1 \\ 1 & 1 \end{bmatrix}$. Next, Picard applied mixed $\sigma_2$ and $I$, which transformed the penny to $\rho_2=(pro) \sigma_2\rho_1 \sigma_2^{\dagger}+(1-pro) I \rho_1 I^{\dagger}=\frac{1}{2}\begin{bmatrix} 1 & 1-2pro \\  1-2pro & 1 \end{bmatrix}$. Finally, Q applied U(*,$\frac{\pi}{2}$) and transformed the penny to $\rho_3= U(*,\frac{\pi}{2}) \rho_2 U(*,\frac{\pi}{2})^{\dagger} =\frac{1}{2}\begin{bmatrix} 1 & a \\ \bar{a} & 1 \end{bmatrix}$.  When measured, the penny collapsed to $|0\rangle$ and to $|1\rangle$ with equal probability, hence both players received expected payoff 0. 

Similar to the strategies in category 1, both $\phi$ of Q's $U1$ and $\theta$ of Q's $U2$ have no impact on the penny's final state $\rho_3$. Figure \ref{case2_q} shows that their values have a wide range and averaged to $\frac{\pi}{2}$ and $\frac{\pi}{3}$. Similarly, Picard's $pro$ has no impact on the penny's final state. Figure \ref{case2_p} shows its values fluctuated between 0 and 1, with an average between 0.4 and 0.6.
 
\subsubsection{Category 3 Quantum Strategies:}
\begin{figure}[!htb]
\begin{minipage}[t]{0.49\linewidth}
\centerline{\includegraphics[width=6.5cm,height=4.0cm]{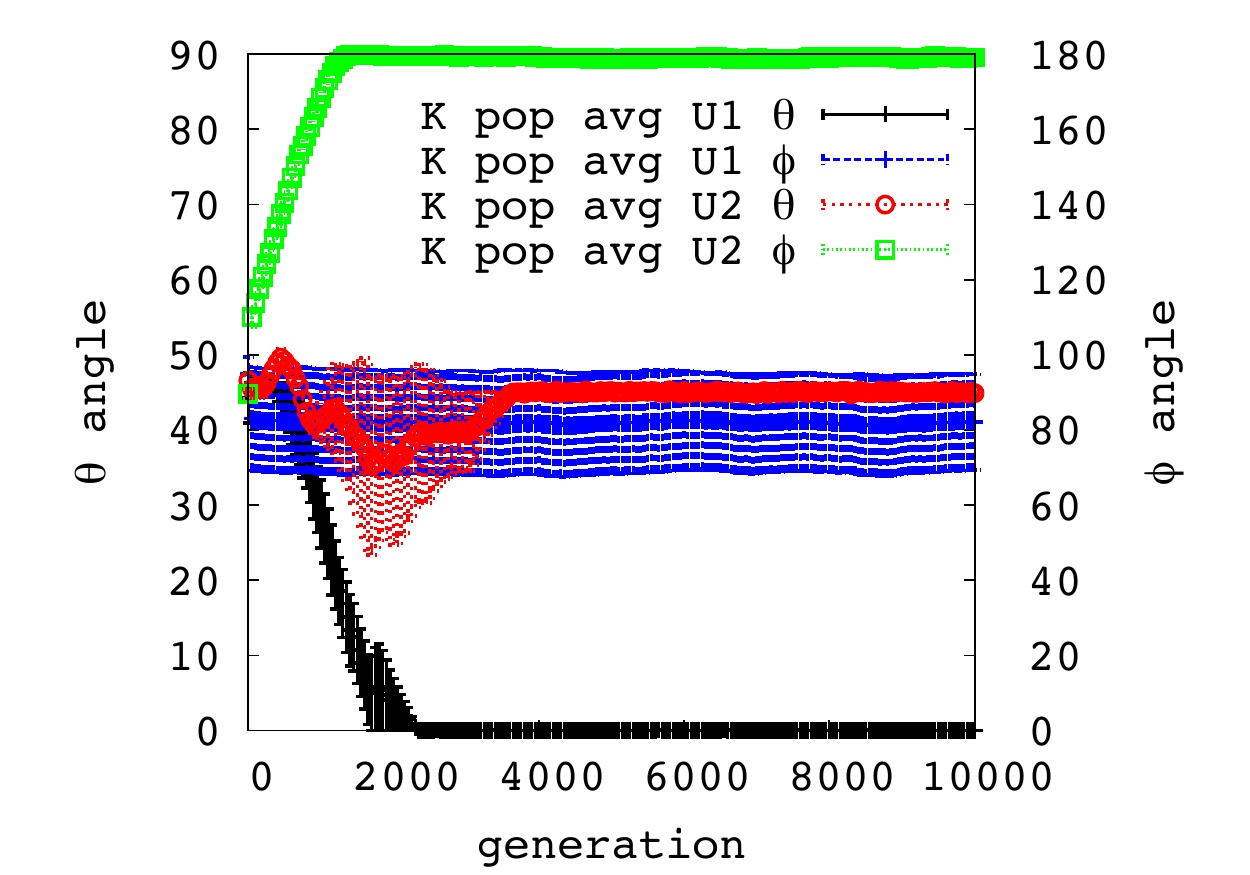}}
\caption{Evolved Q strategies
\label{case3_q}}
\end{minipage}
\begin{minipage}[t]{0.49\linewidth}
\centerline{\includegraphics[width=6.5cm,height=4.0cm]{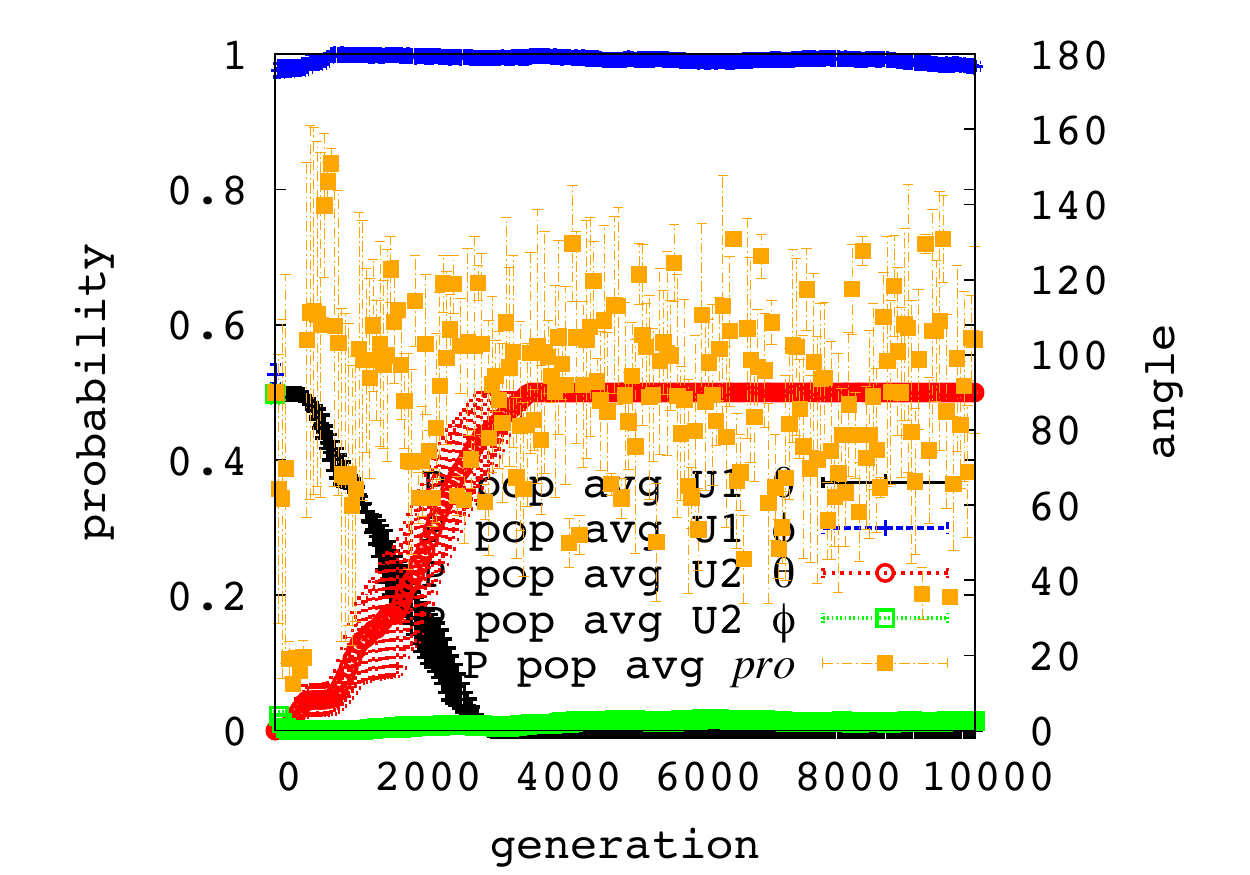}}
\caption{Evolved Picard strategies.
\label{case3_p}}
\end{minipage}
\end{figure}
3 of the 100 simulation runs converged to strategies in category 3, where Q uses U(0, *) and Hadamard to play against Picard's mixed $\sigma_3$ and $\sigma_2$ strategies. With the initial state $\rho_0=\begin{bmatrix} 1 & 0 \\ 0 & 0 \end{bmatrix}$, after Q applied U(0, *), the state of the penny remained the same: $\rho_1=U(0,*)\begin{bmatrix} 1 & 0 \\ 0 & 0 \end{bmatrix}U(0,*)^{\dagger}=\begin{bmatrix} 1 & 0 \\ 0 & 0 \end{bmatrix}$. Next, Picard applied mixed $\sigma_3$ and $\sigma_2$, which transformed the penny to $\rho_2=(pro) \sigma_3 \rho_1 \sigma_3 ^{\dagger}+(1-pro)\sigma_2\rho_1\sigma_2^{\dagger}=\begin{bmatrix} pro & 0 \\  0 & 1-pro \end{bmatrix}$. Finally, Q applied Hadamard and transformed the penny to $\rho_3= H \rho_2 H^{\dagger} =\frac{1}{2}\begin{bmatrix} 1 & 2pro-1 \\ 2pro-1 & 1 \end{bmatrix}$.  When measured, the penny collapsed to $|0\rangle$ and to $|1\rangle$ with equal probability, hence both players received expected payoff 0. 

Similar to the strategies in the two previous categories, $\phi$ of Q's $U1$ and $pro$ of Picard's mixing probability have no impact on the penny's final state. As a result, they did not converge to a particular value. However, their standard errors are much lager than that of the strategies in the two previous categories, because they are averaged over 3 runs, instead of a larger number of simulation runs.


\subsubsection{Category 4 Quantum Strategies:}
\begin{figure}[!htb]
\begin{minipage}[t]{0.49\linewidth}
\centerline{\includegraphics[width=6.5cm,height=4.0cm]{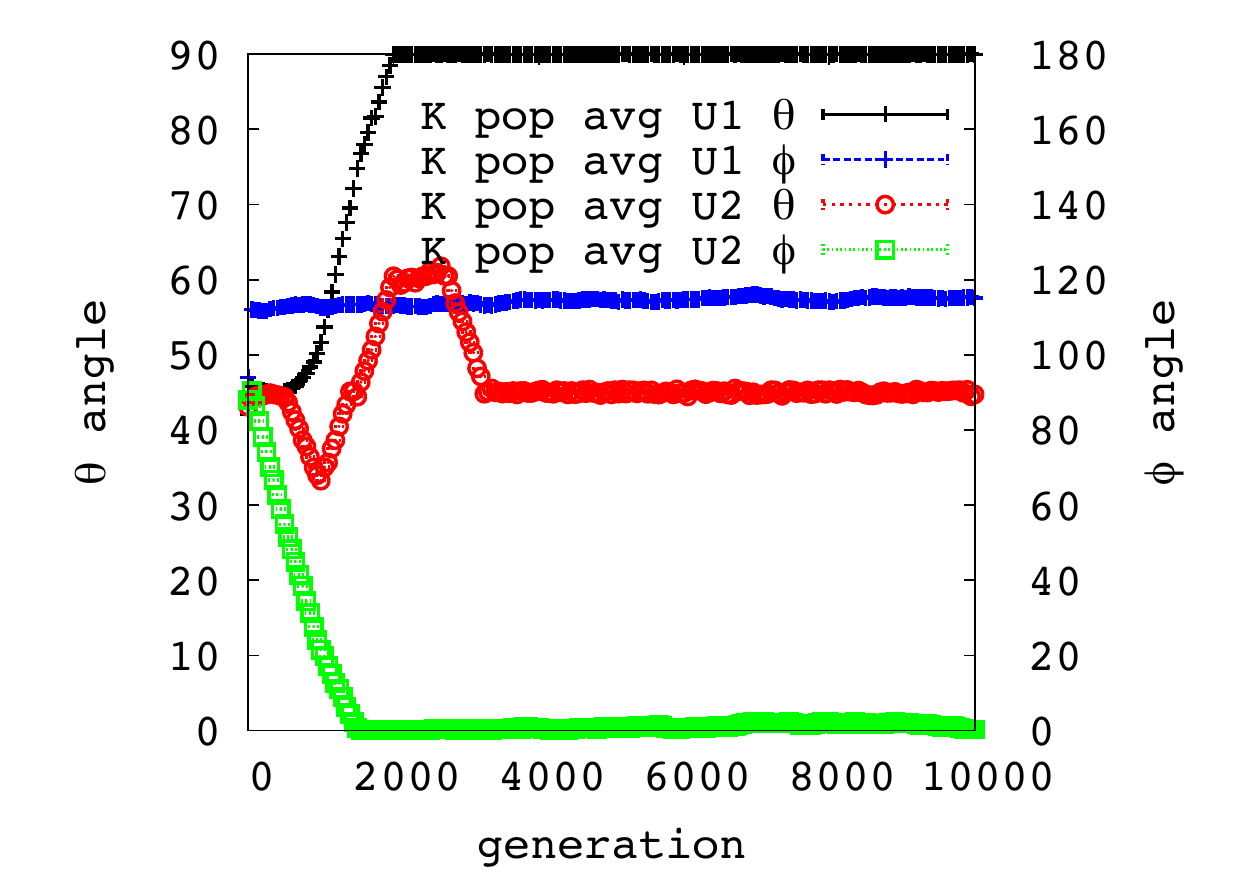}}
\caption{Evolved Q strategies
\label{case4_q}}
\end{minipage}
\begin{minipage}[t]{0.49\linewidth}
\centerline{\includegraphics[width=6.5cm,height=4.0cm]{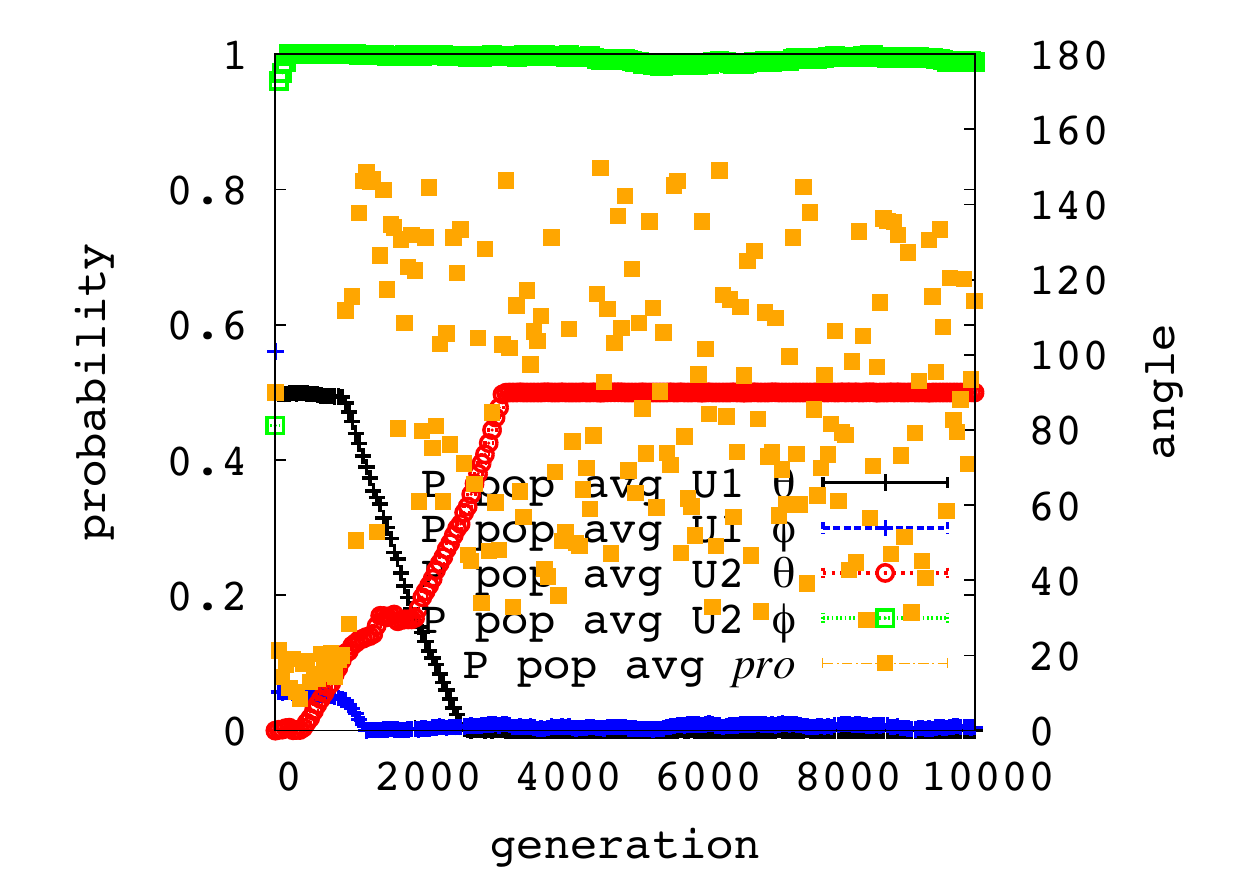}}
\center 
\caption{Evolved Picard strategies.
\label{case4_p}}
\end{minipage}
\end{figure}

Only 1 out of the 100 simulation runs converged to strategies in category 4, where Q used U($\frac{\pi}{2}$,*) and U($\frac{\pi}{4}$,0) to play against Picard's mixed $I$ and $\sigma_1$ strategies. With the initial state $\rho_0=\begin{bmatrix} 1 & 0 \\ 0 & 0 \end{bmatrix}$, after Q applied U($\frac{\pi}{2}$, *), the state of the penny is $\rho_1=U(\frac{\pi}{2},*)\begin{bmatrix} 1 & 0 \\ 0 & 0 \end{bmatrix}U(\frac{\pi}{2},*)^{\dagger}=\begin{bmatrix} 0 & 0 \\ 0 & 1 \end{bmatrix}$.
Next, Picard applied mixed $I$ and $\sigma_1$ strategies to transform the penny to $\rho_2= (pro)I\rho_1I^{\dagger}+(1-pro)\sigma_1\rho_1\sigma_1^{\dagger}=\begin{bmatrix} 1-pro & 0 \\  0 & pro \end{bmatrix}$. 
Finally, Q applied U($\frac{\pi}{4}$,0) which transformed the penny to $\rho_3= U(\frac{\pi}{4},0) \rho_2 U(\frac{\pi}{4},0)^{\dagger}=\frac{1}{2}\begin{bmatrix} 1 & 1-2pro \\ 1-2pro & 1 \end{bmatrix}$.  When measured, the penny collapses to $|0\rangle$ and to $|1\rangle$ with equal probability, hence both players receive expected payoff 0.  Similar to the strategies in the previous three categories, $\phi$ of Q's $U1$ and $pro$ of Picard's mixing probability have no impact on the penny's final state. Therefore, they did not converge to a particular value. 

\subsubsection{Discussion}

Categories 1 and 2 strategies are NE because neither of the two players can change his strategy alone to improve his payoff. However, this is not the case categories 3 and 4 strategies. For example, instead of using U(0,*) and H to force a tie, Q can use U($\frac{\pi}{4}$, *) and U($\frac{\pi}{4}$, 0) to beat Picard's mixed $\sigma_3$ and $\sigma_2$. However, if Q play that strategy, Picard can use the mixed $I$ and $\sigma_1$ strategies to beat Q. Then if Picard use that strategy, Q has another winning strategy U($\frac{\pi}{4}$,*) and H that can beat Picard and win the game. But if Q use that strategy, Picard can use mixed $\sigma_3$ and $\sigma_2$ to beat Q. This loops back to our starting point where Q has a winning strategy U($\frac{\pi}{4}$,*) and U($\frac{\pi}{4}$, 0) to beat PicardÕs mixed $\sigma_3$ and $\sigma_2$.

This circular competition relationship among Q and Picard's winning strategies seems to suggest that there is no equilibrium strategies to settle  those winning strategies. However, under the competitive co-evolution of the EGT model, where both P and K populations are allowed to continuously evolve new strategies to play against the new strategies evolved by the other player, a new set of compromised strategies emerged. The strategies in categories 3 and 4 are able to play against the other player's winning strategies and force a tie. Together, the four categories of quantum strategies form the ES set of this version of the quantum penny flip game.

Compared to simulation 1, where Picard's \emph{mixed-two classical strategies} in the ES set were not able to change the state of the penny transformed by Q's quantum strategy, hence lost the game every time, in this simulation, the \emph{mixed-two quantum strategies} in the ES set allowed Picard to always force a tie. In other words, quantum strategies have benefited Picard in this version of the quantum penny flip game. 

\section{Concluding Remarks}\label{conclude}
Classical game theory is a mature science that is frequently applied to analyze conflicts that arise during decision making  in economics and social sciences. With the recent development of quantum information processing, many classical games have been quantized to investigate the game dynamics under the influence of quantum mechanics. However, there has not been work applying evolutionary game theory (EGT) models to investigate quantum game theory.  This work proposed and developed an EGT model to investigate the quantum penny flip games. In particular, we used the model to conduct a series of simulations where a population of mixed classical strategies from the ES set of the game were invaded by quantum strategies.  The results of our investigation are very encouraging.

First, we found that when only one of the two players' mixed classical strategies were invaded, the results were different. In one case, due to the interference phenomenon of superposition,  quantum strategies provided more payoff, hence successfully replaced the mixed classical strategies in the ES set. In  the other case, the mixed classical strategies were able to sustain the invasion of quantum strategies and remained in the ES set. Secondly, when both players' mixed classical strategies were invaded by quantum strategies, a new quantum ES set emerged. The strategies in the quantum ES set give both players payoff 0, which is the same as the payoff of the strategies in the mixed classical ES set of this game.

With the established EGT framework, we will continue our investigation of mixed quantum ES set in the quantum penny flip game. In particular, we will increase the number of quantum strategies used by each player to identify other quantum ES sets in this game \cite{miszczak_etal}. We are also interested in applying the developed methodology to study other quantum games.

 \end{document}